\documentclass[journal]{IEEEtran}

\usepackage{amsmath,amsfonts,amssymb}
\usepackage{algorithmic}
\usepackage{algorithm}
\usepackage{array}
\usepackage[caption=false,font=normalsize,labelfont=sf,textfont=sf]{subfig}
\usepackage{textcomp}
\usepackage{stfloats}
\usepackage{url}
\usepackage{verbatim}
\usepackage{graphicx}
\usepackage{cite}

\usepackage{xcolor}
\usepackage{siunitx}
\usepackage{booktabs}

\usepackage{xurl}
\usepackage{tabularx}
\usepackage{diagbox}
\usepackage{makecell}

\usepackage[printonlyused,nohyperlinks]{acronym}

\renewcommand{\baselinestretch}{0.955}

\usepackage{eurosym}

\usepackage{tikz}

\usepackage{bbding}
\usepackage{pifont}

\usepackage{multirow}

\usepackage{adjustbox}

\usepackage{scalerel}
\usepackage{tikz}
\usetikzlibrary{svg.path}

\definecolor{orcidlogocol}{HTML}{A6CE39}
\tikzset{
	orcidlogo/.pic={
		\fill[orcidlogocol] svg{M256,128c0,70.7-57.3,128-128,128C57.3,256,0,198.7,0,128C0,57.3,57.3,0,128,0C198.7,0,256,57.3,256,128z};
		\fill[white] svg{M86.3,186.2H70.9V79.1h15.4v48.4V186.2z}
		svg{M108.9,79.1h41.6c39.6,0,57,28.3,57,53.6c0,27.5-21.5,53.6-56.8,53.6h-41.8V79.1z M124.3,172.4h24.5c34.9,0,42.9-26.5,42.9-39.7c0-21.5-13.7-39.7-43.7-39.7h-23.7V172.4z}
		svg{M88.7,56.8c0,5.5-4.5,10.1-10.1,10.1c-5.6,0-10.1-4.6-10.1-10.1c0-5.6,4.5-10.1,10.1-10.1C84.2,46.7,88.7,51.3,88.7,56.8z};
	}
}

\newcommand\orcidicon[1]{\href{https://orcid.org/#1}{\mbox{\scalerel*{
				\begin{tikzpicture}[yscale=-1,transform shape]
					\pic{orcidlogo};
				\end{tikzpicture}
			}{|}}}}

\usepackage{hyperref}

\title{AI Safety Assurance for Automated Vehicles: \\A Survey on Research, Standardization, Regulation}

\author{Lars Ullrich \orcidicon{0009-0001-8166-3118} \IEEEauthorrefmark{2}, Michael Buchholz \orcidicon{0000-0001-5973-0794} \IEEEauthorrefmark{3}, Klaus Dietmayer \orcidicon{0000-0002-1651-014X} \IEEEauthorrefmark{3},~\IEEEmembership{Senior Member,~IEEE,} \\and Knut Graichen \orcidicon{0000-0003-2865-8093} \IEEEauthorrefmark{2},~\IEEEmembership{Senior Member,~IEEE}
	\thanks{*This research is accomplished within the project ”AUTOtech.agil” (FKZ 01IS22088Y, FKZ 01IS22088W). We acknowledge the financial support for the project by the Federal Ministry of Education and Research of Germany (BMBF).\\Corresponding author: {\tt\footnotesize lars.ullrich@fau.de}}
	\thanks{\IEEEauthorrefmark{2}Chair of Automatic Control, Friedrich-Alexander-Universität Erlangen-Nürnberg (FAU), Cauerstraße 7, 91058 Erlangen, Germany {\tt\footnotesize \{lars.ullrich, knut.graichen\}@fau.de}}
	\thanks{\IEEEauthorrefmark{3}Institute of Measurement, Control and Microtechnology, Ulm University, Albert-Einstein-Allee 41, 89081 Ulm, Germany {\tt\footnotesize \{michael.buchholz, klaus.dietmayer\}@uni-ulm.de}}
	\thanks{Manuscript received July 31, 2024; revised October 03, 2024, October 29, 2024, November 04, 2024; accepted November 10, 2024.}}

\usepackage{hologo}
\usepackage{listings}
\begin{document}
	
\pagestyle{empty}
\twocolumn[
\begin{@twocolumnfalse}
	\Huge {IEEE copyright notice} \\ \\
	\large {\copyright\ 2024 IEEE. Personal use of this material is permitted. Permission from IEEE must be obtained for all other uses, in any current or future media, including reprinting/republishing this material for advertising or promotional purposes, creating new collective works, for resale or redistribution to servers or lists, or reuse of any copyrighted component of this work in other works.} \\ \\
	
	{\Large Published in \emph{IEEE Transactions on Intelligent Vehicles}, 15 November 2024.} \\ \\
	
	Cite as:
	
	\vspace{0.1cm}
	\noindent\fbox{%
		\parbox{\textwidth}{%
			L.~Ullrich, M.~Buchholz, K.~Dietmayer, and K.~Graichen, ''AI Safety Assurance for Automated Vehicles: A Survey on Research, Standardization, Regulation,''
			in \emph{IEEE Transactions on Intelligent Vehicles}, 15 November 2024, pp. 1--19, doi: 10.1109/TIV.2024.3496797.
		}%
	}
	\vspace{2cm}
	
\end{@twocolumnfalse}
]

\noindent\begin{minipage}{\textwidth}
	
\hologo{BibTeX}:
\footnotesize
\begin{lstlisting}[frame=single]
@article{ullrich2024ai,
       title={AI Safety Assurance for Automated Vehicles: A Survey on Research, Standardization, Regulation},
       author={Ullrich, Lars and Buchholz, Michael and Dietmayer, Klaus and Graichen, Knut},
       journal={IEEE Transactions on Intelligent Vehicles},
       year={2024},
       pages={1--19},
       doi={10.1109/TIV.2024.3496797},
       publisher={IEEE}
}
\end{lstlisting}
\end{minipage}

\maketitle
\setcounter{page}{1}

\begin{abstract}\label{00_Abstract}
	Assuring safety of artificial intelligence (AI) applied to safety-critical systems is of paramount importance. Especially since research in the field of automated driving shows that AI is able to outperform classical approaches, to handle higher complexities, and to reach new levels of autonomy. At the same time, the safety assurance required for the use of AI in such safety-critical systems is still not in place. Due to the dynamic and far-reaching nature of the technology, research on safeguarding AI is being conducted in parallel to AI standardization and regulation. The parallel progress necessitates simultaneous consideration in order to carry out targeted research and development of AI systems in the context of automated driving. Therefore, in contrast to existing surveys that focus primarily on research aspects, this paper considers research, standardization and regulation in a concise way. Accordingly, the survey takes into account the interdependencies arising from the triplet of research, standardization and regulation in a  forward-looking perspective and anticipates and discusses open questions and possible future directions. In this way, the survey ultimately serves to provide researchers and safety experts with a compact, holistic perspective that discusses the current status, emerging trends, and possible future developments. 
\end{abstract}


\section{Introduction}\label{1}
Autonomous vehicles are well-known safety-critical systems that are expected to have a significant impact on a wide range of societal and economic domains \cite{crayton2017autonomous, montgomery2018america}. In this context, the transformation towards autonomous mobility is based on technological progress \cite{chen2023milestones}. In particular, technological advancements in artificial intelligence (AI) are promising towards achieving higher levels of autonomy \cite{grigorescu2020survey}. Thereby, the characteristic of the black-box nature offers, e.g., the merit of mapping the system and behavior without extracting and modeling the explicit relationships, but instead emulating the input/output behavior based on data. However, this also implies that the behavior is no longer clearly evident and the existing analytical methods for so-called white-box models, especially for safety and reliability argumentation, cannot be applied directly \cite{flammini2020safety, lazarus2020runtime}.

With regard to AI safety, a far-reaching description of prevailing challenges was presented back in 2016 \cite{amodei2016concrete}. Even though the technology of AI has developed considerably since then, the challenges of AI safety addressed back then are predominantly still present. At the same time, the rapid further development raises the question of what is defined as AI in this paper. In this respect, we adhere to the Organisation for Economic Co-operation and Development's (OECD) definition, which is quite far-reaching but also widely accepted. In short, AI is a machine-based system that influences the environment to achieve specific objectives by perceiving it based on input data, abstracting these perceptions into models and generating outputs. Thereby, AI systems operate with varying degrees of autonomy \cite{OECDpub}. 

In general, the provability of AI safety is referred to as AI safety assurance. In this paper, we understand AI safety assurance according to the definition of \cite{batarseh2021survey}, which refers to a process throughout the AI lifecycle to ensure that AI outputs are valid, verified, data-driven, trustworthy, and explainable, but also ethical, unbiased, and fair. In this context, reference should also be made to a comprehensive systematic literature review on AI safety assurance \cite{neto2022safety}. At the same time, the question of the novelty of this survey arises. In fact, there is a large number of AI surveys \cite{habli2020artificial, kuznietsov2024explainable}, but with very different objectives. Although there is a series of AI safety-related surveys conducted within the field of automated driving (AD) \cite{perez2024artificial, waschle2022review, nascimento2019systematic}, these mostly go less into the depth of AI safety assurance, while at the same time focusing on specific perspectives and therefore allow only limited conclusions to be drawn about AI safety assurance methodology in general and for the future.

Thus, a key objective is to facilitate further considerations, such as in \cite{diaz2023connecting}, since technology is developing rapidly and standards and regulations are evolving in parallel. However, while \cite{diaz2023connecting} addresses the trustworthiness of AI in general and other publications such as \cite{dafoe2018ai} and \cite{anderljung2023frontier} explore AI governance and regulation, a further consideration tailored to the application of automated driving is missing, but is crucial as automated driving is a high-risk system with far-reaching implications and pre-existing application-specific constraints. 

The main contribution of this paper is the joint consideration of the application of automated driving and the technology of AI with a focus on safety assurance along the triad of research, standardization and regulation. Thereby, the application of automated driving refers to the automated vehicle itself and its functionalities, not to the extended vehicle or the entire intelligent transportation system. In this way, the survey enables the consideration of interdependencies of multiple factors within a delimited, specific application area, which is important to achieve a forward-looking perspective with an appropriate depth. Our study thus provides various stakeholders, from functional researchers and developers to safety experts and regulators, with a holistic view and enables the recognition of possible developments. In short, the key findings of the survey suggest that a shift towards data-based safety assurance is necessary to address the characteristics of AI systems and the resulting challenges.

For the selection of references throughout the paper, we relied on extensive experience with safety-critical automotive systems, complemented by a targeted review of relevant standards, laws and scientific literature. We used recognized publication databases and applied expert knowledge to identify the most relevant references to ensure broad coverage of the field. Moreover, the reference choice is briefly described along the survey. 

The paper is structured as follows: to present the survey at a glance, the key findings, characteristics, limitations, contributions, etc. are laid out in Section \ref{KeyIdeas}. Thereafter, the current state of the art and the future prospects of AI safety research are outlined in Section \ref{SafetyResearch_chapter2}. The AI safety related standardization along the application of autonomous driving is presented in Section \ref{SafetyStandardization_chapter3}. The AI regulatory landscape, with a particular focus on the EU, the USA, and China as well as other selected countries, is examined in Section \ref{SafetyRegulation_chapter4}. Finally, open questions and potential solutions are addressed in Section \ref{sol}, while Section \ref{SafetySummary_chapter5} provides summary and conclusion on research, standardization, and regulation.
\section{Survey at a Glance}\label{KeyIdeas}
In this section, the most essential elements and aspects of the survey are briefly presented before the paper goes into detail. 

\textit{Key Findings:} Recent surveys show a growing awareness and demand for AI safety assurance from various perspectives, from research over standardization up to regulation. At the same time, all affected areas are facing significant challenges and have a number of unanswered questions.

\textit{Core Features:} The three core features of AI safety assurance across different perspectives are, i) the increasing awareness of data dependency, ii) the recognition of the need for lifecycle considerations, and iii) the need for safety abstraction, e.g., through functional safety or Safety of the Intended Functionality (SOTIF).

\textit{General Limitations:} Despite increasing AI safety assurance efforts, the methods currently in use are unable to address the challenges inherent in AI systems. Furthermore, AI safety standardization in the automotive sector is lagging behind general AI standardization in terms of data dependency awareness and lifecycle consideration, while general AI standardization is also still underway. In addition, AI regulations around the globe pose additional hurdles for products intended for different markets, as they are very heterogeneous.

\textit{Main Contributions:} This study contributes by examining the three pillars of research, standardization, and regulation of AI safety assurance for automated vehicles, identifying interdependencies and highlighting the need for approaches to safety assurance in terms of methods, standards and regulation. In particular, the need for a data-driven AI safety assurance approach that takes into account i) the inherent challenges, ii) the lifecycle considerations, and iii) the necessary agnosticism to accommodate future innovations. But it also calls for open innovation and close collaboration on open standards and regulations that serve as non-legally binding guidelines that encourage new pathways and ensure continuous improvement.

\textit{New Insights:} The shift towards data-driven AI safety assurance corresponds to the underlying change from mathematically explicit systems to data-based implicit systems, and thus represents the corresponding natural counterpart evolution to system design in the area of system analysis and safety assurance. 

\textit{Knowledge Synthesis:} The current state of AI safety assurance in research, standardization, and regulation addresses essential core features, but is subject to general limitations. In this context, this survey facilitates the understanding of different perspectives and developments and their interdependencies. It shows that the transition to data-based AI safety assurance over the entire lifecycle with an appropriate level of abstraction is promising in many respects, but that the possible solution is also accompanied by open questions that a systematic transformation of existing methods can address.
\section{AI Safety Research}\label{SafetyResearch_chapter2}
This section deals with the current status of AI systems and corresponding safety research. Furthermore, future perspectives in this area are outlined. Finally, the section concludes with a discussion.

\subsection{Current State of AI Systems \& Safety Research}
In recent years, AI safety research has become increasingly important. Nevertheless, there is a large number of open research questions \cite{amodei2016concrete}, some of which are fundamental, while at the same time promising new AI methods are being introduced. A wide-ranging examination of current state of the art AI safety assurance, including open research questions, is presented in \cite{neto2022safety}. It appears that current safety research primarily covers simple AI systems.

In comparison, modern AI systems are mostly based on complex architectures. For example, in the area of trajectory prediction of other road users, different combinations of common model architectures, such as convolutional neural networks (CNN), recurrent neural networks (RNN), Transformers, and recently even Large Language Models (LLM), are considered \cite{gu2021densetnt, varadarajan2022multipath, shi2022motion, lan2024traj}. The choice of overall architecture depends on the data to be processed and the design decisions. One approach, for instance, is to use several pre-trained general-purpose models, also known as backbones, which are then tailored to the respective task through application-specific training. For example, CNN backbone models \cite{liu2020cbnet, wang2020cspnet} could be used to process image data such as maps from a bird's eye view, while RNN and transformer backbones \cite{wang2022predrnn, wang2021pyramid} could be used to process temporal data such as the histories of road users as part of trajectory prediction. Besides individual submodules for the respective data processing steps, another approach for predicting trajectories is to use a comprehensive advanced model for the task as a whole. For example, the exploration of targeted transformers \cite{shi2022motion} or LLMs \cite{lan2024traj}. Furthermore, this consideration can also be applied to the entire automated driving stack, ultimately leading to an End-to-End architecture \cite{chen2023end, cui2023drivellm}. Regardless of the granularity of the AI (sub)systems used, the overall complexity is increasing. At the same time, new promising methodological approaches are being explored \cite{assran2023self}, thus the overall complexity is likely to increase further in the future. In the realm of such complex systems, formal proofs of AI safety cannot be expected at this stage. Specifically, as the behavior of AI systems also depends on data, as a large number of researchers have already noted \cite{burton2021safety, gauerhof2020assuring, salay2019improving}. For instance, data used at development time affects the weights during training. And data at runtime influences the system behavior. To give an example, the behavior of the long short-term memory cells of recurrent networks \cite{hochreiter1997long} or the multi-head self-attention of transformers \cite{vaswani2017attention} depend on operational data.\\
On the one hand, this data dependency further complicates the safety guarantee process \cite{neto2022safety}. On the other hand, a bottom-up safety assurance approach, where individual backbone models are pre-validated and then fine-tuned for a specific task, proves to be impractical and insufficient. This is because fine-tuning modifies the weights of the model and thus re-specifies the system behavior, rendering all preliminary proofs or validations obsolete \cite{amodei2016concrete, chen2019catastrophic, qi2023fine}. As a result, new proofs are required to ensure the safety of the finalized AI unit, taking into account both task-specific development and operational data \cite{neto2022safety}. This is exemplified in the analysis process for traffic sign detection and recognition systems using neural networks \cite{chaghazardi2023explainable, berghoff2021towards}. Consequently, it can be concluded that regardless of the chosen safety approach, a data-based or data-dependent safety assurance approach is mandatory.

A formal safety assurance approach that accounts for data dependency might be theoretically possible. However, given that AI systems are designed to map highly complex, non-trivial relationships, the notion of a general, formal, closed-form solution appears contradictory. Both the definition and feasibility of achieving such a goal are open to question, despite its desirability.

This becomes clear when examining the challenges for AI safety that arise from the change in data characteristics between training and operation \cite{zhou2022domain}. While initial research has been conducted in this area, e.g., on the safety assurance implications of input data \cite{burton2021safety, gauerhof2020assuring, salay2019improving}, it remains unclear how factors such as data biases, dataset shifts, concept shifts, out-of-domain data or data quality will specifically impact safety. The effects of redundant training data, such as correlated data or redundant AI models, e.g., ensembles, on safety have also not yet been clarified \cite{neto2022safety}.

While in this context, it is desirable to elaborate formal correlations for AI safety assurance, data-driven performance evaluation highlights the factors that require investigation today. This is because performance degradation and feature implications are observed during evaluations. Rather than focusing on formal correlations, current procedures emphasize the development of new requirements and methods to address emerging issues. This performance-oriented AI development follows a cycle of exploration, observation, and mitigation, and serves as a key stimulus for ongoing research. Consequently, data-driven evaluation and observation provide insights and enable conclusions about reliability, robustness, and resilience of AI systems and drive emerging methods such as \cite{pmlr-v37-romera-paredes15, vinyals2016matching, wang2020generalizing, hochreiter2001learning, vanschoren2019meta} that address existing performance limitations. Accordingly, it must be questioned to what extent the intended decidedly formal statements about interrelationships and influencing factors represent an appropriate goal.
 
\subsection{Prospects on AI Systems \& Safety Research}
In general, assurance is based on defining requirements and assumptions and verifying the system’s correct functionality at each point in time given these assumptions \cite{burton2021safety, burton2023addressing, iso26262, iso21448}. However, AI systems handle complex tasks using data-based and implicit methods to emulate desired system behaviors. As a result, system assumptions are embedded in both the architecture and the training data, while assumptions about the admissible application domain are reflected in the test data. To properly verify these assumptions, a data-driven approach is essential, as recognized by \cite{perez2024artificial, rabe2021development}. The shift from analytically explicit assumptions to data-driven implicit ones requires an equivalent shift in verification and validation methods toward data-centric assurance approaches. In summary, we refer to the associated approach to safety assurance, which builds on implicit assumptions embedded within data and the data-based verification and validation of assumptions through to data-based system analysis, as data-driven AI safety assurance. Although this is a comparatively broad formulation, the specification and refinement of this definition will remain a subject of future research.

The data-based assurance approach outlined here aligns to some extent with current practices, though it is not always explicitly communicated. Research has already demonstrated that, for certain architectures and use cases, operational data can differ significantly from development data \cite{stacke2020measuring}. To verify the underlying assumptions, methods like out-of-distribution (OoD) detection have been developed \cite{liu2020energy, ren2019likelihood}. For instance, during deployment, OoD detection identifies deviations between operational and training data distributions. In other words, the implicit assumptions embedded in the training data, when compared to the operational data, can be continuously monitored. Such assumption-checking mechanisms enable the implementation of safeguarding measures \cite{shafaei2018uncertainty, wilson2023safe}.

However, there are use cases where changes in assumptions, such as shifts in data distribution, are inherent. To address these cases and maintain correct functionality, various methods have been developed, including zero-shot \cite{pmlr-v37-romera-paredes15, 10.1145/3293318}, one-shot \cite{fei2006one, vinyals2016matching}, few-shot \cite{wang2020generalizing, kadam2020review}, and meta-learning techniques \cite{hochreiter2001learning, finn2017model, vanschoren2019meta}. It seems that in performance-oriented development, data-driven approaches have already gained unconscious traction. Adopting similar methods for validation appears to be a purposeful step. However, these approaches must evolve into a systematic, process-oriented methodology worth of verification and validation.

At first glance, the shift to data-based assurance may appear disruptive and radical. However, a closer examination reveals that both safety assurance and performance-oriented development tackle similar challenges and address comparable open questions. For instance, \cite{neto2022safety} highlights the investigation of redundant data and structures as a key safety assurance research question. In fact, several models in trajectory prediction \cite{varadarajan2022multipath, shi2022motion} extract redundant latent data from various sources and use different models to validate and fuse this information, often through attention mechanisms. Although this approach within trajectory prediction is mainly performance-driven, it inherently uncovers assurance requirements. Additionally, functional safety standards like \cite{iso26262, iso21448} already blend performance-oriented aspects with safety considerations, demonstrating that this transition is more gradual and integrated than it might initially appear.

The data-driven repetitive process of exploration, observation, and mitigation provides a way to validate implicit assumptions in the architecture, training, and test data. If observations confirm the assumptions, no changes are needed. If not, mitigation is required. This can be done by either narrowing the scope to ensure the assumptions hold or adjusting the methodology so that new assumptions are met. The choice depends on the application, but continuous monitoring is advisable in safety-critical systems to detect and respond to invalid assumptions that could lead to unpredictable behavior.

While this data-driven approach cannot be directly applied in safety-critical systems in the real-world, it can be addressed through simulation-based development and subsequent transfer, e.g. through transfer learning \cite{pan2009survey}. For instance, training and evaluation can initially be performed in simulation, which enables more efficient data generation and reuse \cite{yurtsever2020survey, sankaranarayanan2018learning}. In particular, the ability to generate hazardous scenario data safely and efficiently by means of simulations is a significant advantage and offers the possibility of improving methodological robustness \cite{ding2020learning, wang2021advsim, ding2023survey} without real-world dangers. In this context, the concept of imaginative intelligence discussed in \cite{wang2024does} represents a promising research direction. Nevertheless, it is clear that the use of simulators and the associated gap between simulation and reality entails a certain risk \cite{bargman2024methodological}. In particular, it is essential to ensure that no inaccuracies are introduced that could undermine credibility and trustworthiness \cite{neto2022safety}. In this context, especially the imaginative intelligence exhibits open questions and dangers besides the possibilities \cite{wang2024does}. As a result of existing drawbacks, a gradual transition back to the real world is required from a safety assurance perspective, in which iterative improvements are made to both the simulation and the system in order to minimize the gap and thus the risk.

Therefore, the transition to real systems is indispensable for validating assumptions and reaching maturity. This mirrors findings in developmental psychology, which emphasize that experimentation is key to learning and development \cite{gopnik2004theory}. Real-world validation is necessary for system approval and ongoing improvements. Accordingly, data-based verification in real applications is necessary to achieve approval and continuous improvement. However, this does not mean that systems are used in the real world without existing validation or approval. Rather, this corresponds to the approach that is already being used in practice in the field of autonomous driving. Namely, building on simulation-based validation in the real world to collect data and the gradual transfer of more responsibility and scope for action under the supervision of trained staff, like Waymo Limited Liability Company (LCC) or Cruise LCC are doing already today.

\subsection{Discussion}
A key advantage of moving to data-driven safety assurance is the architecture- and technology-independent nature of the methodology. This approach allows for the rapid safeguarding
of various complex and novel systems, without requiring safety practitioners to delve into the AI architectures and methodologies. As a result, proposed paradigm shift towards data-driven safety assurance enables a long-term, transferable and practical approach to AI safety assurance. Furthermore, this independence from AI architectures and methods is also leveraged in the latest AI Safety Integrity Level (AI-SIL) approach \cite{diemert2023safety}. However, data-driven safety assurance places greater emphasis on the methodology rather than AI-SIL classification. Nevertheless, the joint development of these approaches represents a promising direction for future research.

Moreover, while we agree with \cite{burton2023closing} in their analysis, we propose a different solution. Specifically, we challenge the notion that narrow AI is the only viable solution, given recent technological advancements. General AI, in our view, holds potential for handling corner cases more effectively at higher levels of autonomy. Furthermore, end-to-end learning, particularly modular end-to-end learning, has demonstrated the potential for enhanced performance \cite{levine2016end, eysenbach2022mismatched, leong2022bridging, chen2023end}, which directly contributes to improving safety.

We also share the recognition of the need for robustness and resilience in AI systems \cite{burton2023closing}. However, we find the proposed solution of adaptive online learning critical, particularly in terms of safety and fairness. Taking automated driving as an example, safety would depend on the individual use of the vehicle. However, the behavior of an autonomous vehicle should be consistent across the entire fleet. Moreover, the real world is mutable, meaning that system requirements will change over time. 

To address this, data must be collected continuously, and system admissibility regularly checked and renewed. This process, however, is resource-intensive, requiring significant computational and memory capacities. Therefore, we advocate for regular offline learning and updates, similar to periodic reviews of vehicle systems, like Germany’s periodic technical inspection \cite{tuvsud2024}. While this provides an outlook specific to automated driving, the principle can be applied to other domains requiring ongoing assurance of AI systems.

Finally, it should be noted that technology agnosticism can offer a great advantage on the long run. The use of graphical processing units for AI systems \cite{cirecsan2010deep, schmidhuber2015deep} has significantly changed the development and methodology of AI. In this context, the relevance of a hardware agnostic functional and data-oriented methodology that could be quickly transferable becomes evident.
\section{AI Safety Standardization}\label{SafetyStandardization_chapter3}
The safety standardization for methods and techniques is usually based on the corresponding safety research as well as on existing and known safety approaches that are transferred. However, as already mentioned, there is a large number of open research questions in the field of AI safety research. In addition, the previous analysis shows the need for data-based or data-driven assurance regardless of the methodology. However, data-driven assurance is still relatively uncommon in safety standardization. Nevertheless, the significant improvement in the performance of AI methods is urging the parallel development of standards that ultimately govern their application. The following provides an overview of current and developing standards related to AI and autonomous driving. 

\subsection{AI Safety Standardization from a Methodological Point of View}
In the last years, a number of standards have been published in the field of information technology and artificial intelligence. Table \ref{tab:standards_subsection_a} provides an overview and selection of relevant standards for AI safety assurance and related topics. These encompass a wide array of themes including safety, security, trust, ethics, \& societal concerns, as well as robustness, data, software, and AI lifecycle \& process management. This comprehensive coverage indicates that standardization in the realm of general AI safety assurance and related topics is already taking a holistic approach.

As it turns out, safety in relation to AI is addressed in \cite{iso5469}, with a particular focus on functional safety, which provides agnosticism about methodologies and hardware setups. While such a safety approach is also discussed in previous Section II, the corresponding analysis of AI safety research goes further and demands the linkage or centering of functional safety in relation to data. 

At the same time, it becomes apparent that the relevance of data and the consideration of lifecycles within information technology and AI standardization is recognized and taken into account. This can be seen both in the recently published standards and in the standards still under development presented in Table \ref{tab:standards_subsection_a}. This demonstrates that a functional and data-oriented methodology seems to be the logical next step from both AI safety assurance research and standardization.

Moreover, in addition to the standardization of AI safety assurance in general form, application-oriented standardizations are emerging in parallel. In particular, safety-critical cyber-physical systems are dealing with the topic. Two well-known representatives are the aerospace and the automotive sector. In line with the underlying focus of this paper, the automotive sector is examined in more detail subsequently.

\begin{table*}
	\centering
	\caption{General overview of standards for AI safety assurance and related topics.}
	\begin{tabularx}{\linewidth}{l l c X c c}
		\toprule
		\textbf{Group} &
		\textbf{Standard} &
		\textbf{Category}$^{1}$ & 
		\textbf{Short Title} & 
		\textbf{Status}$^{2}$ &
		\textbf{Source} \\
		\midrule
		Safety, Security, & ISO/IEC TR 5469:2024 & IT, AI &
		Functional safety and AI systems & PS & \cite{iso5469} \\
		\multirow{1}{2.2cm}{Trust, Ethics, \& \\ Societal Concern} & ISO/TR 22100-5:2021 & SM &  Relationship with ISO 12100, Part 5: Implications of artificial intelligence machine learning & PS & \cite{iso22100} \\
		& ISO 13849-2:2012 & SM & Safety-related parts of control systems, Part 2: Validation & PS & \cite{iso13849_2} \\
		& ISO/IEC TR 24028:2020 & IT, AI&  
		Overview of trustworthiness in artificial intelligence & PS & \cite{iso24028} \\
		& ISO/IEC TS 8200:2024 & IT, AI & Controllability of automated artificial intelligence systems & PS & \cite{iso8200} \\
		& ISO/IEC 23894:2023 & IT, AI & Guidance on risk management & PS & \cite{iso23894} \\
		& ISO/IEC TR 24027:2021 & IT, AI &
		Bias in AI systems and AI aided decision making & PS & \cite{iso24027}  \\	
		& ISO/IEC CD 27090 & IT, SP & Guidance for addressing security threats and failures in artificial intelligence systems & UD &\cite{iso27090} \\
		& ISO/IEC TR 27563:2023 & IT, SP & Security and privacy in artificial intelligence use cases — Best practices & PS & \cite{iso27563} \\
		& ISO/IEC TR 24368:2022 & IT, AI & Overview of ethical and societal concerns & PS & \cite{iso24368} \\
		\midrule
		Robustness & ISO/IEC TR 24029-1:2021 & IT, AI & Assessment of the robustness of neural networks, Part 1: Overview & PS &  \cite{iso24029-1:2021}  \\
		& ISO/IEC 24029-2:2023 & IT, AI &
		Assessment of the robustness of neural networks, Part 2: Methodology for the use of formal methods & PS & \cite{iso24029-2}  \\
		& DIN SPEC 92001-2:2020-12 & (IT, AI) & 
		Künstliche Intelligenz - Life Cycle Prozesse und Qualitätsanforderungen - Teil 2: Robustheit & PS & \cite{din92001-12} \\
		\midrule
		Data & ISO/IEC 5259-1:2024 & IT, AI & Data quality for analytics and machine learning, \quad \quad \quad  \quad \quad Part 1: Overview, terminology, and examples & PS & \cite{iso5259-1} \\
		& ISO/IEC FDIS 5259-2 & IT, AI &
		Part 2: Data quality measures & UD & \cite{iso5259-2}  \\
		& ISO/IEC 5259-3:2024 & IT, AI &
		Part 3: Data quality management requirements and guidelines & PS & \cite{iso5259-3}  \\
		& ISO/IEC 5259-4 & IT, AI &
		Part 4: Data quality process framework & UP & \cite{iso5259-4} \\
		& ISO/IEC DIS 5259-5 & IT, AI &
		Part 5: Data quality governance & UD & \cite{iso5259-5}   \\
		& ISO/IEC CD TR 5259-6 & IT, AI &
		Part 6: Visualization framework for data quality & UD & \cite{iso5259-6} \\
		& ISO/IEC AWI TR 42103 & IT, AI &
		Overview of synthetic data in the context of AI systems & UD & \cite{iso42103} \\
		\midrule
		Software & ISO/IEC/IEEE 12207:2017 & IT, SS & Software life cycle processes & PS & \cite{iso12207} \\
		& ISO/IEC/IEEE 15026-2:2022 & IT, SS & Systems and software assurance, Part 2: Assurance case & PS & \cite{iso15026} \\
		& ISO/IEC/IEEE 15026-4:2021 & IT, SS & Systems and software assurance, Part 4: Assurance in the life cycle & PS & \cite{iso15026_4} \\
		& ISO/IEC TS 25058:2024 & IT, AI & Systems and software Quality Requirements and Evaluation, Guidance for quality evaluation of artificial intelligence systems & PS & \cite{iso25058} \\
		& ISO/IEC 25059:2023 & IT, AI &	Software engineering — Systems and software Quality Requirements and Evaluation — Quality model for AI systems & PS & \cite{iso25059} \\
		\midrule
		AI Lifecycle \& & ISO/IEC 5338:2023 & IT, AI &
		AI system life cycle processes & PS & \cite{iso5338} \\
		Process Management & ISO/IEC 8183:2023 & IT, AI &
		Data life cycle framework & PS & \cite{iso8183}  \\
		& ISO/IEC TS 33061:2021 & IT, SS & Process assessment model for software life cycle processes & PS & \cite{iso33061-2021} \\
		& ISO/IEC 24668:2022 & IT, AI &
		Process management framework for big data analytics & PS &\cite{iso24668}  \\
		& ISO/IEC AWI TS 17847 & IT, AI &
		Verification and validation analysis of AI systems & UD & \cite{iso17847} \\
		& ISO/IEC DIS 42006 & IT, AI &
		Requirements for bodies providing audit and certification of artificial intelligence management systems & UD & \cite{iso42006}  \\
		& ISO/IEC AWI TR 42106 & &
		Overview of differentiated benchmarking of AI system quality characteristics & UD & \cite{iso42106}  \\
		\bottomrule
	\end{tabularx}
	\label{tab:standards_subsection_a}
	\begin{minipage}{\textwidth}
	\scriptsize
	\begin{itemize}
	\item[$^{1}$] IT (Information Technology), AI (Artificial Intelligence), SM (Safety \& Machinery), SP (Security \& Privacy Protection), SS (Software \& Systems Engineering), ES (Electrical, Electronic \& General System), RV (Road Vehicles), VS (Vehicle Systems), RTS (Road Traffic Safety Systems), ITS (Intelligent Transport Systems), ST (Safety \& Testing)
	\item[$^{2}$] PS (Published), UP (Under Publication), UD (Under Development) \\
	\end{itemize}
	\end{minipage}
\end{table*}

\subsection{AI Safety Standardization from an Automotive Perspective}
In the automotive sector, there is a large number of standards. Given the focus on automated vehicles, the following overview of AI safety standardization is specifically tailored to address safety-critical driving functions, as well as assistance and automation systems related to the vehicle itself. By contrast, the extended vehicle, e.g., intelligent transportation systems, and communication technologies such as Vehicle-to-Everything (V2X) or Internet of Vehicles (IoV) are outside the scope of this analysis. Readers interested in these areas are referred to following the publications \cite{machardy2018v2x, zhao2018vehicular, limbasiya2019iovcom, contreras2017internet, taslimasa2023security, buchholz2021handling, storck2020survey, zeadally2020vehicular} and the standards \cite{ISO_20077, ISO_14813, ISO_21217, etsi302_665, etsi302_637, etsi303_613, etsi103_723}. Nevertheless, the following analysis remains centered on the vehicle itself and its automation systems. In accordance a selection of related standards for safety in automated and AI-based intelligent vehicle systems is represented in Table \ref{tab:standards_section_b}. 

As indicated by Table \ref{tab:standards_section_b}, functional safety occupies a special position in the automotive sector. In particular, ISO 26262 \cite{iso26262}, which evolved from IEC 61508 \cite{iso61508}, represents its backbone. Building on this, there is a number of standards that deal with safety, from Functional Safety (FuSa) \cite{iso61508, iso26262, iso8926} to SOTIF \cite{iso21448}. Morover, standards that address the safety of automated driving systems \cite{iso5083} and artificial intelligence \cite{iso8800} are particularly relevant. As it can be seen, artificial intelligence in the automotive sector, just as in general standardization, is increasingly considered \cite{iso27090, iso8800}. 

Another central aspect of standardization in the automotive sector addresses scenario-based testing and evaluation, as seen in Table \ref{tab:standards_section_b}. Beyond that, big data and AI \cite{iso12786} are also covered by recently published standards and standards that are currently under development. 

In addition to the previously mentioned safety and scenario-based testing, the operational design domain (ODD) plays a crucial role in general testing within the automotive sector \cite{iso17720}. Furthermore, for automated driving systems, a recently published standard on the evaluation of autonomous products is essential for ensuring advanced systems safety and effectiveness \cite{isoUL4600}. Another closely related topic is security, as security, which is concerned with the protection of systems from malicious attacks, is an important prerequisite for achieving system safety. For instance, a security threat represents a risk that could lead a system to potentially unsafe behavior, if not mitigated. This illustrates that safety and security are closely intertwined, which is in relation to AI systems also outlined in \cite{neto2022safety}. Moreover, while \cite{bertino2021ai} provides a general discussion on the relationship between AI and security, the cybersecurity of connected automated vehicles is particularly examined, e.g., within publications such as \cite{han2023secure, guo2023sustainability}. Furthermore, some articles specifically examine the relationship between safety and security in automated driving \cite{dutta2018security}, as well as the analysis of adversarial attacks on AI and corresponding detection systems to mitigate these threats \cite{teng2023paid}.

Overall, in the context of AI-based systems in the automotive sector, security primarily focuses on cybersecurity throughout the development lifecycle, including design, verification, validation \cite{iso4804} and general engineering practices \cite{iso21434, isoSAEJ3061}. As AI systems are exposed to the environment and process all inputs, safety and security are more closely linked in the respective AI context. This suggests that data will also become more important in terms of security. 

Beyond that, the software plays a substantial role within the automotive sector. Thereby, the updatability of software \cite{iso24089-2023, iso25090} and the software lifecycle \cite{iso33061-2021} in general are of particular relevant. Moreover, over-the-air updates \cite{iso24935} and service-oriented diagnostics \cite{iso17978} should be mentioned in this regard, which reflect the general development towards software-defined vehicles and service-oriented architectures. An overview of the software related standards is provided in the respective group within Table \ref{tab:standards_section_b}. 

Alongside the more general standards, there is also a number of specific standards in the automotive sector. These encompass, for e.g, specific driver assistance systems such as highway chauffeur \cite{iso23792-1, iso23792-2} or automated valet parking systems \cite{iso23374-1, iso23374-2}. Moreover, there are also standards covering function-specifics such as minimum risk maneuvers \cite{iso23793-1}. A selection of standards is presented under the advanced driver-assistance systems (ADAS) group in Table \ref{tab:standards_section_b}.

Compared to general AI safety standardization, far less explicit importance is attached to the subject of data. While data is implicitly involved, especially in test scenarios, it is not generally accounted for in the safety assurance within the automotive sector. As noted by \cite{koopman2019safety}, neither functional safety nor SOTIF sufficiently address the unique safety challenges posed by AI-based systems in automotive applications. While we agree with the analysis and the view that goal-based safety cases, like those outlined in ANSI/UL 4600 \cite{isoUL4600}, are beneficial, we also share the insight from \cite{diemert2023safety} that a fundamental change is necessary, including the need for a new SIL definition, such as AI-SIL \cite{diemert2023safety}. While the shift in SIL definition is still at the conceptual stage, a broader methodological shift is demanded. Specifically, as suggested by the AI safety research analysis, a shift towards a data-driven assurance process.

Additionally, as \cite{koopman2019safety} also highlights, updating overall processes towards iterative methods is essential to achieve the necessary flexibility for AI systems. While these data-driven, iterative processes and adapted safety methodologies are becoming more prominent in AI standardization efforts, e.g., Table \ref{tab:standards_subsection_a}, subsection "AI Lifecycle \& Process Management", they have not yet been adopted in the automotive sector.

\begin{table*}[!t]
	\centering
	\caption{Automated vehicles centered overview of standards for AI safety assurance and related topics.}
	\begin{tabularx}{\linewidth}{l l c X c c}
		\toprule
		\textbf{Group} &
		\textbf{Standard} &
		\textbf{Category}$^{1}$ & 
		\textbf{Short Title} & 
		\textbf{Status}$^{2}$ &
		\textbf{Source} \\
		\midrule
		\multirow{1}{1.0cm}{Safety,\\Security, \&} & IEC 61508 & (ES) &
		Functional safety of electrical/electronic/programmable electronic safety-related systems & PS & \cite{iso61508} \\
		\& Ethics & ISO 26262:2018 & RV, ES &
		Functional safety & PS & \cite{iso26262} \\
		& ISO 21448:2022 & &
		Safety of the intended functionality & PS & \cite{iso21448} \\
		& ISO/TR 9968:2023 & RV, ES & Functional safety — Application to generic rechargeable energy storage systems for new energy vehicle & PS & \cite{iso9968} \\
		& ISO/TR 9839:2023 & RV, ES & Application of predictive maintenance to hardware with ISO 26262-5 & PS & \cite{iso9839} \\
		& PAS 1881:2020 & (RV) &
		Assuring safety for autonomous vehicle trials and testing –
		Specification & PS & \cite{iso1881}  \\
		& ISO/TR 4804:2020 & RV &
		Safety and cybersecurity for automated driving systems — Design, verification and validation & PS & \cite{iso4804} \\	
		& SAE J3061 & (IT / RV) & Cybersecurity Guidebook for Cyber-Physical Vehicle Systems & PS & \cite{isoSAEJ3061} \\
		& ISO/SAE 21434:2021 & RV, ES &  Cybersecurity engineering & PS & \cite{iso21434} \\
		& ISO/PAS 8926:2024 & RV, ES &
		Functional safety — Use of pre-existing software architectural elements & PS & \cite{iso8926}\\
		& ISO/CD TS 5083 & RV, ES & Safety for automated driving systems — Design, verification and validation & UD & \cite{iso5083}  \\		
		& SAE J3206:2021 &(RV, ES)& Taxonomy and Definition of Safety Principles for Automated Driving System & PS & \cite{seaj3206} \\
		& ISO/DPAS 8800 & RV, ES &
		Safety and artificial intelligence & UD & \cite{iso8800} \\
		& ISO 39003:2023 & RTS & Guidance on ethical considerations relating to safety for autonomous vehicles & PS & \cite{iso39003} \\
		\midrule
		Testing \& & ISO 34501:2022 & RV, VS & Test scenarios for automated driving systems — Vocabulary & PS & \cite{iso34501} \\
		Evaluation & ISO 34502:2022 & RV, VS & Test scenarios for automated driving systems — Scenario based safety evaluation framework & PS & \cite{iso34502-2022} \\
		& ISO 34503:2023 & RV, VS &
		Test scenarios for automated driving systems — Specification for operational design domain & PS & \cite{iso34503-2023}  \\
		& ISO 34504:2024 & RV, VS &
		Test scenarios for automated driving systems — Scenario categorization & PS & \cite{iso34504} \\
		& ISO/DIS 34505 & RV, VS &
		Test scenarios for automated driving systems — Scenario evaluation and test case generation & UD & \cite{iso34505} \\
		&ISO/TS 22133:2023 & RV, VS & Test object monitoring and control for active safety and automated/autonomous vehicle testing — Functional requirements, specifications and communication protocol & PS & \cite{iso22133} \\
		& ANSI/UL 4600 & (RV) &
		Standard for Evaluation of Autonomous Products & PS & \cite{isoUL4600} \\
		& ISO/CD TR 12786 & ITS &
		Big data and artificial intelligence supporting intelligent transport systems - Use cases & UD & \cite{iso12786} \\
		& ISO/AWI TR 17720 & ITS & Operational Design Domain Boundary and Attribute Awareness for an Automated Driving System & UD & \cite{iso17720} \\
		\midrule
		Data & ISO/TR 21718:2019 & ITS & Spatio-temporal data dictionary for cooperative ITS and automated driving systems 2.0 & PS & \cite{iso21718} \\
		& ISO 20524-2:2020 & ITS & Geographic Data Files, Part 2: Map data used in automated driving systems, Cooperative ITS, and multi-modal transport & PS & \cite{iso20524} \\
		& ISO/TS 22726:2023 & ITS & Dynamic data and map database specification for connected and automated driving system applications & PS & \cite{iso22726} \\ 
		\midrule
		Software & ISO 24089:2023 & RV, ES & Software update engineering & PS & \cite{iso24089-2023} \\
		& ISO/AWI PAS 25090 & RV, ES & Software Update engineering - vehicle configuration information & UD & \cite{iso25090} \\
		& ISO/AWI TR 24935 & RV, ES & Software Update over the air using mobile cellular network & UD & \cite{iso24935} \\
		& ISO/AWI 17978  & RV & Service-oriented vehicle diagnostics & UD & \cite{iso17978} \\
		\midrule
		ADAS & ISO 21717:2018 & ITS & Partially Automated In-Lane Driving Systems — Performance requirements and test procedures & PS & \cite{iso21717} \\
		& ISO/TS 23792-1:2023 & ITS & Motorway chauffeur systems, Part 1: Framework and general requirements & PS & \cite{iso23792-1} \\
		& ISO/CD 23792-2  & ITS & Motorway chauffeur systems, Part 2: Requirements and test procedures for discretionary lane change & UD & \cite{iso23792-2} \\
		& ISO/AWI 19484 & ITS & Highly Automated Motorway Chauffeur Systems & UD & \cite{iso19484} \\
		& ISO 23374-1:2023 & ITS &
		Automated valet parking systems, Part 1: System framework, requirements for automated driving and for communications interface & PS & \cite{iso23374-1} \\
		& ISO/TS 23374-2:2023 & ITS & Automated valet parking systems, Part 2: Security integration for type 3 AVP & PS & \cite{iso23374-2}\\
		& ISO/TR 5255-2:2023 & ITS & Low-speed automated driving system service, Part 2: Gap analysis & PS & \cite{iso5255} \\
		& ISO 20900:2023 & ITS & Partially-automated parking systems — Performance requirements and test procedures & PS & \cite{iso20900} \\ 
		& ISO 23375:2023 & ITS &  Collision evasive lateral manoeuvre systems — Requirements and test procedures & PS & \cite{iso23375} \\
		& ISO 23793-1  & ITS & Minimal risk manoeuvre for automated driving, Part 1: Framework, straight-stop and in-lane stop & UP & \cite{iso23793-1} \\
		\bottomrule
	\end{tabularx}
	\label{tab:standards_section_b}
		\begin{minipage}{\textwidth}
		\scriptsize
		\begin{itemize}
			\item[$^{1}$] IT (Information Technology), AI (Artificial Intelligence), SM (Safety \& Machinery), SP (Security \& Privacy Protection), SS (Software \& Systems Engineering), ES (Electrical, Electronic \& General System), RV (Road Vehicles), VS (Vehicle Systems), RTS (Road Traffic Safety Systems), ITS (Intelligent Transport Systems), ST (Safety \& Testing)
			\item[$^{2}$] PS (Published), UP (Under Publication), UD (Under Development) \\
		\end{itemize}
	\end{minipage}
\end{table*}

\subsection{Emerging AI Safety Standardization: Collaborative Efforts}
While the previous subsections have mainly referred to current and recently published standards and to a limited extent to standards under development, this subsection is mainly dedicated to standards that are work in progress. An overview of technical committees, subcommittees, working groups, and standard committees in the intersection ofAI safety assurance and autonomous driving is presented in Table \ref{tab:standards_section_c}. 

Overall, as it can be seen from Table \ref{tab:standards_section_c}, a large number of standards are under development. In more detail, Table \ref{tab:standards_section_c} provides an overview about 102 standards that are work in progress within the field of AI safety assurance for automated driving. Thereby, the largest portion, around 25\% emerge from the Joint Technical Committee (JTC) 1, Subcommittee (SC) 42, artificial intelligence and around 17 \% from the Technical Committee (TC) 204, intelligent transportation systems.

Overall, as shown in Table \ref{tab:standards_section_c}, a significant number of standards are currently under development. Specifically, Table \ref{tab:standards_section_c} highlights 102 ongoing standards in the field of AI safety assurance for automated driving. The largest share, approximately 25\%, originates from the Joint Technical Committee (JTC) 1, Subcommittee (SC) 42 on artificial intelligence, around 17\% from the Technical Committee (TC) 204 on intelligent transportation systems, and about 10\% from the JTC 1, SC 7 on software and systems engineering. In general, within the automotive sector, a number of standards are still focusing on individual ADAS functionalities, with increasing automation and expansion of the ODD. At the same time, however, standards are also being developed that deal with end-to-end AI concepts. Moreover, it is apparent that most standards under development within the field of AI address safety, robustness, data, and the lifecycle taking on a central role. 

Furthermore, reviewing the working groups and related publications is useful for anticipating the content of future standards. Therefore, in the intersection of AI safety assurance and autonomous driving, reference is made at this point to \cite{burton2020mind, burton2023closing, burton2023addressing}. These publications specifically address the complexity of the system and the resulting challenges, namely uncertainties, insufficiencies, and gaps. 

Beyond the technical uncertainties and performance insufficiencies that are prevalent in application-oriented research, uncertainties in specification and assurance are addressed. These are usually neglected in function-oriented research, but have a decisive impact on the overall safety. For this reason, all contributors should be aware of these sources of uncertainty. The publications also recommend handling by means of functional safety based on ISO 21448 \cite{iso21448}, namely SOTIF. In this context, \cite{burton2023addressing} addresses reasoning about input data and assigns greater importance to data for the first time. 

Moreover, a detailed analysis of the standards that are currently work in progress would also be desirable. Due to the fact that the standards are not publicly available in this state and are only available for internet use, a more detailed analysis of them is not possible at the present time and therefore recourse to related publications is most expedient. However, in general, it can be seen from the titles and efforts, that in the area of AI safety research, the importance of data continues to have a stronger weighting and in the area of automotive standards is only considered more pointedly, in accordance with the title, twice.

\begingroup
\renewcommand{\arraystretch}{0.97}  
\renewcommand{\baselinestretch}{0.92}

\begin{table*}[!t]
	\centering
	\caption{Overview of work in progress standards for AI safety assurance and automated vehicles.}
	\begin{tabularx}{\linewidth}{ l l  c  X  l  }
		\toprule
		\textbf{Org.} &
		\textbf{Division}$^{1}$ &
		\textbf{Category}$^{2}$ & 
		\textbf{Short Title} &
		\textbf{Work in Progress}\\
		\midrule
		\textbf{ISO}&\textbf{TC 22} & \textbf{RV} & \textbf{Road vehicles} & \\
		\cmidrule{2-5}
		&TC 22/SC 32	&RV, ES& Electrical and electronic components and general system aspects & \multirow{3}{6cm}{ISO/CD TS 5083, ISO/SAE CD PAS 8475, ISO/SAE AWI TR 8477, ISO/DPAS 8800, ISO 24089:2023/Amd 1, ISO/AWI TR 24935, ISO/AWI PAS 25090} \\
		& TC 22/SC 32/WG 8  &RV, ES& 	Functional safety & \\
		&TC 22/SC 32/WG 11 &RV, ES&	Cybersecurity & \\
		& TC 22/SC 32/WG 12  &RV, ES&	Software update & \\
		& TC 22/SC 32/WG 13  &RV, ES&	Safety for driving automation systems \\
		& TC 22/SC 32/WG 14  &RV, ES&	Safety and Artificial Intelligence & \\
		\cmidrule{2-5}
		& TC 22/SC 33 &RV, VS & Vehicle dynamics, chassis components and driving automation systems testing & \multirow{4}{6cm}{ISO/CD 3888-2, ISO/AWI 11010-2, ISO/AWI PAS 11585-2, ISO/WD TS 19206-9, ISO/AWI 22133, ISO/AWI 25135, ISO/AWI PAS 25321, ISO/DIS 34505} \\
		&TC 22/SC 33/WG 3  &RV, VS&	Driver assistance and active safety functions & \\
		&TC 22/SC 33/WG 9  &RV, VS&Test scenarios of automated driving systems	& \\
		&TC 22/SC 33/WG 11  &RV, VS&	Simulation & \\
		&TC 22/SC 33/WG 16  &RV, VS&	Active Safety test equipment &	
		 \\
		\cmidrule{2-5}
		&TC 22/SC 36 &RV, ST& Safety and impact testing & \multirow{2}{6cm}{ISO/AWI TS 23520, ISO/DTS 21934-2, ISO/DTR 12353-4, ISO/CD TS 4654} \\
		&TC 22/SC 36/WG 4  &RV, ST&	Virtual testing & \\
		&TC 22/SC 36/WG 7 &RV, ST& Traffic accident analysis methodology & \\
		\cmidrule{2-5}
		&\textbf{TC 199} &\textbf{SM}& \textbf{Safety of machinery} & \multirow{1}{6cm}{ISO/CD 13849-2, ISO/AWI TR 13849-3} \\
		&TC 199/WG 8 &SM&Safe Control Systems & \\
		\cmidrule{2-5}
		&\textbf{TC 204} & \textbf{ITS}&\textbf{Intelligent transport systems} & \multirow{1}{6cm}{ISO/CD 12768-1, ISO/AWI 12768-2, ISO/CD TR 12786, ISO/AWI 15622, ISO/AWI 17387, ISO/AWI TR 17720, ISO/DIS 18750, ISO/AWI 19237, ISO/AWI 19484, ISO/AWI 21717, ISO/AWI TS 22726-1, ISO/DTS 22726-2, ISO/AWI 23375, ISO/AWI 23792-1, ISO/CD 23792-2, ISO 23793-1, ISO/CD TR 24856} \\
		&TC 204/WG 14 &ITS& Vehicle/roadway warning and control systems	& \\
		&TC 204/WG 20 &ITS& Big Data and Artificial Intelligence supporting ITS	& \\ \\ \\  
		\cmidrule{2-5}
		&\textbf{TC 241}&\textbf{RTS}& \textbf{Road traffic safety management systems} & \multirow{1}{6cm}{ISO/WD 39004} \\
		&TC 241/WG 6 &RTS&Guidance on ethical considerations relating to autonomous vehicles & \\
		\cmidrule{1-5}
		\textbf{ISO/IEC} &\textbf{JTC 1}&\textbf{IT}&\textbf{Information technology}& \\
		\cmidrule{2-5}
		&JTC 1/SC 7&IT, SS&Software and systems engineering& \multirow{1}{6cm}{ISO/IEC CD 9837-1, ISO/IEC/IEEE DIS 15026-1, ISO/IEC AWI TS 22864, ISO/IEC/IEEE CD 23612, ISO/IEC/IEEE CD 24748-4.2, ISO/IEC PRF 25040, ISO/IEC/IEEE CD 25070, ISO/IEC/IEEE FDIS 29119-5, ISO/IEC CD TR 29119-8.2, ISO/IEC AWI TS 33062}\\
		&JTC 1/SC 7/AHG 9&IT, SS&AI-based software development&\\
		&JTC 1/SC 7/WG 7&IT, SS&Life cycle management&\\
		&JTC 1/SC 7/WG 10&IT, SS&Process assessment&\\
		&JTC 1/SC 7/WG 26&IT, SS&Software testing&\\
		&JTC 1/SC 7/WG 30&IT, SS&Systems resilience&\\
		&JTC 1/SC 42/JWG 2&IT, SS&Testing of AI-based systems&\\
		\cmidrule{2-5}
		
		&JTC 1/SC 27&IT, SP&Information security, cybersecurity and privacy protection& \multirow{1}{6cm}{ISO/IEC CD 5181, ISO/IEC AWI TS 5689, ISO/IEC DIS 15408-2, ISO/IEC DIS 15408-3, ISO/IEC AWI 27045, 
			ISO/IEC CD 27090, ISO/IEC WD 27091.2} \\
		&JTC 1/SC 27/AG 2&IT, SP&Trustworthiness&\\
		&JTC 1/SC 27/AHG 3&IT, SP&Security and privacy in AI and Big Data&\\
		&JTC 1/SC 27/JWG 6&IT, SP&Cybersecurity requirements and evaluation activities for connected vehicle devices&\\
		&JTC 1/SC 27/WG 1&IT, SP&Information security management systems&\\
		&JTC 1/SC 27/WG 3&IT, SP&Security evaluation, testing and specification& \\
		\cmidrule{2-5}
		&JTC 1/SC 42/JWG 2&IT, AI&Testing of AI-based systems& \multirow{1}{6cm}{ISO/IEC AWI 4213, ISO/IEC FDIS 5259-2, ISO/IEC 5259-4, ISO/IEC DIS 5259-5, ISO/IEC CD TR 5259-6, ISO/IEC CD TS 6254, ISO/IEC DTS 12791.2, ISO/IEC DIS 12792, ISO/IEC AWI TS 17847, ISO/IEC AWI TS 22440-1, ISO/IEC AWI TS 22440-2, ISO/IEC AWI TS 22440-3, ISO/IEC AWI TS 22443, ISO/IEC 22989:2022/AWI Amd 1, ISO/IEC 23053:2022/AWI Amd 1, ISO/IEC AWI 24029-3, ISO/IEC AWI 25059, ISO/IEC AWI TS 29119-11, ISO/IEC DIS 42005, ISO/IEC DIS 42006, ISO/IEC AWI 42102, ISO/IEC AWI TR 42103, ISO/IEC AWI 42105, ISO/IEC AWI TR 42106, ISO/IEC AWI TR 42109}\\
		&JTC 1/SC 42/JWG 4&IT, AI&Functional safety and AI systems&\\
		&JTC 1/SC 42/WG 1&IT, AI&Foundational standards&\\
		&JTC 1/SC 42/WG 2&IT, AI&Data&\\
		&JTC 1/SC 42/WG 3&IT, AI&Trustworthiness&\\
		&JTC 1/SC 42/WG 4&IT, AI&Use cases and applications&\\
		&JTC 1/SC 42/WG 5&IT, AI&Computational approaches and computational characteristics of AI systems&\\ \\ \\ \\ \\ \\
		\midrule
		\textbf{IEEE}& FSSC&(ES)& Functional Safety Standards Committee&	P2851.1\\
		\cmidrule{2-5}
		&C/S2ESC&(IT, SS)&Software \& Systems Engineering Standards Committee & P1228 \\
		\cmidrule{2-5}
		&VTS/AVSC&(RV, VS)&Vehicular Technology Society - Automated Vehicles Standards Committee&		P2846a, P2979, P3116, P3321\\
		\cmidrule{2-5}
		&ITSS/SC&(ITS)&Intelligent Transportation Systems Society Standards Commitee&P3412, P3344\\ 
		\cmidrule{2-5}
		&C/AISC&(IT, AI)&Artificial Intelligence Standards Committee&\multirow{1}{6cm}{P2863, P3123, P3127, P3157, P3142, P3187, P3193, P3198, P3342, P3395, P3423, P3426, P3378}\\ \\
		\bottomrule
	\end{tabularx}
	\label{tab:standards_section_c}
	\begin{minipage}{\textwidth}
		\scriptsize
		\begin{itemize}
			\item[$^{1}$] TC (Technical Committee), JTC (Joint Technical Committee), SC (Subcommittee), WG (Working Group), JWG (Joint Working Group), AG (Advisory Group)
			\item[$^{2}$] IT (Information Technology), AI (Artificial Intelligence), SM (Safety \& Machinery), SP (Security \& Privacy Protection), SS (Software \& Systems Engineering), ES (Electrical, Electronic \& General System), RV (Road Vehicles), VS (Vehicle Systems), RTS (Road Traffic Safety Systems), ITS (Intelligent Transport Systems), ST (Safety \& Testing)
		\end{itemize}
	\end{minipage}
\end{table*}
\endgroup

\subsection{Discussion}
In general, this section served to provide a bundled presentation of the overall activities within the area of standardization. It can be noted that data is increasingly being taken into account, particularly in the case of AI standardization, while in the case of autonomous driving, the focus is mainly on specific scenario data. 

Regarding the automotive sector, there is still a lack of specific standardization with a stronger focus on data and the lifecycle of AI systems. Therefore, engineers in the automotive sector should consider appropriate AI standards along the available standards in their domain. However, in the long term, application-specific standards that consider data and the AI lifecycle could be expected, also in line with emerging standards. While application-specific standards sometimes do not add value, they do in the context of AI safety and the application of automated driving. Especially since not every AI is used in a cyber-physical, safety-critical system that is introduced on a large scale. In addition, there is a large number of existing standards, e.g. in the area of hardware and electronics, which must also be taken into account in the automotive sector. The standards must therefore be aligned and harmonized. Application-specific standards are therefore desirable and beneficial as they will enable more efficient development and approval in the long term. 

Overall, the trend towards functional safety, which represents an opportunity to address heterogeneous, complex, and emergent systems due to the agnostic nature, is becoming apparent. However, the systematic fusion of data-based processes with functional safety is not yet part of current standardization. Neither in the field of AI nor in the automotive sector. Accordingly, this results in a general open research question with regard to AI safety assurance processes and associated standardization.
\section{AI Safety Regulation}\label{SafetyRegulation_chapter4}
As technologies come into use and unfold their effect on individuals and/or society, the need for legal consideration arises. Since AI is undoubtedly a powerful technology that is currently being used and will continue to be used even much more intensively in various areas and forms in the future, a legal consideration is appropriate. In recent years, the legal treatment of AI has been addressed at various scales and stages. However, a global approach does not exist. Significant differences in approaches and perspectives exist among key economic and technological regions, therefore a global approach is not expected shortly. Furthermore, in line with the purpose of this survey, the regulation of automated driving is also covered. The brief overview of current regulations and efforts focuses on a selection of important AI and automated driving developing regions and relevant markets, as technology has a push and hand markets a pull effect. Accordingly, the focus was placed on the EU, the US, and China. 

\subsection{European Union (EU)}
The EU is pursuing the path of a central law for artificial intelligence, the EU AI Act (AIA). In April 2021, the European Commission presented an initial proposal for an AI liability directive, an AIA, and an Annex to the AIA in the form of a harmonized legal framework \cite{europeancommissionaiact}. The Commission advocates a technology-neutral definition of AI systems and a risk-based approach to requirements and obligations. More specifically, it distinguishes between unacceptable, high, low, and minimal risk. On this basis, the proposal includes prohibitions for unacceptable risk, strict requirements and obligations for high risk, and relatively lenient transparency obligations for low and minimal risk. The annex to the AIA states that safety-critical AI applications in the real world, such as autonomous driving, should be assigned to the high-risk category.

In December 2022, the European Council adopted a common position of the EU member states, proposing amendments to the Commission's proposal \cite{europeancouncilaiact}. For example, the Council is in favor of a more restrictive definition of AI systems, an exclusion of AI systems for military or national security purposes, and the extension of the ban on social scoring systems by private individuals.

In June 2023, the EU Parliament voted on its position \cite{europarlaiact}. The most important adjustments include several key changes. This includes the definition of AI systems, specifically the use of the definition provided by the OECD. Furthermore, biometric identification systems are entirely prohibited, including the previous exception for terrorist attacks. Additionally, generative AI is explicitly considered, along with specific requirements. Most importantly in this survey, the classification criteria for high-risk systems are tightened, and the competencies of national authorities are strengthened. Beyond that, another crucial adjustment is the promotion of research and development of free and open-source AI through wide-ranging exemptions from compliance with AI law. The most important adjustments include the use of the OECD definition for AI systems, the entire prohibition of biometric identification systems and thus also of the previous execution cases such as terrorist attacks, the explicit consideration of generative AI including specific requirements, the tightening of the classification criterion for high-risk systems, the strengthening of the competences of national authorities, and the promotion of research and development of free and open-source AI through wide-ranging exemptions from compliance with AI law. 

In December 2023, the Council and the Parliament reached a provisional political agreement on the legal text of the EU's AIA. Numerous proposed amendments were taken into account besides the revision and clarification of existing aspects and the incorporation of new aspects. Among other things, the agreement covers the OECD definition, the restriction of the use of remote biometric identification systems by law enforcement authorities, the revision of the governance system, the refinement of obligations for high-risk AI systems, the explicit addressing of general-purpose AI systems, and the expansion of regulatory sandboxes as measures to promote innovation within start-ups or small and medium-sized enterprises (SMEs). To promote innovation and preserve scientific freedom, research and development activities have also been excluded from the scope of the regulation. In addition, systems for military, defense, or national security purposes are also excluded from the scope. With regard to high-risk systems, the clarification of obligations, in particular the mandatory impact assessment with regard to fundamental rights, should be mentioned. Finally, the draft of the EU AIA, whose structure overview is provided in Table \ref{tab:general_structure_regulation}, was adopted by the parliamentary committees on March 13, 2024. While the ban on AI systems with an unacceptable risk will come into force after six months, the code of conduct will apply after nine months and AI systems for general purposes will be regulated after 12 months, the majority of the legislation incl. Annex III will come into force after 24 months and the and the final part, e.g. obligations for high-risk AI systems under Annex II (Article 85), after 36 months. Note that annexes are supplements that are an integral part of the Act, as they contain essential technical details and information that are difficult to consider in the main body.

\begin{table}
	\centering
	\caption{General structure overview of the European Commission's AI legislative proposal \cite{eu_parliament_2024corr}.}
	\begin{tabular}{l p{0.52\linewidth} l}
		\toprule
		Chapter 1 & General provisions & (Art. 1-4) \\
		Chapter 2 & Prohibited AI practices & (Art. 5) \\
		Chapter 3 & High-risk AI systems & (Art. 6-49) \\
		Chapter 4 & Transparency obligations for providers and deployers of certain AI systems & (Art. 50) \\
		Chapter 5 & General-purpose AI models & (Art. 51-56) \\
		Chapter 6 & Measures in support of innovation  & (Art. 57-63) \\
		Chapter 7 & Governance  & (Art. 64-70) \\
		Chapter 8 & EU database for high-risk AI systems & (Art. 71) \\
		Chapter 9 & Post-market monitoring, information sharing and market surveillance & (Art. 72-94) \\
		Chapter 10 & Codes of conduct and guidelines & (Art.95-96) \\
		Chapter 11 & Delegation of power and committee procedure & (Art. 97-98) \\
		Chapter 12 & Penalties & (Art. 99-101) \\
		Chapter 13 & Final provisions & (Art. 102-113) \\
		Annexes & Technical specs. and add. requirements & (Annex I-IX) \\
		\bottomrule
	\end{tabular}
	\label{tab:general_structure_regulation}
\end{table}

As shown in Table \ref{tab:general_structure_regulation}, the EU AIA covers a wide range of topics. Given that the focus of this survey is on AI and automated driving, which is classified as high-risk AI according to the EU AIA, Chapters 3 and 8 are particularly relevant. However, Chapter 9 is also significant in relation to system operation. A more detailed overview of the relevant articles is provided in Table \ref{tab:regulation}. The selection of these articles is based on a careful, experience-based assessment considering which aspects are most relevant in the development of AI-based systems for automated driving throughout the development process but also related to standardization. Without going into detail, the considerable granularity and wide-ranging treatment of relevant aspects of governance, design, development, testing, verification, validation, traceability, certification, deployment, auditing, post-market surveillance, market surveillance, conformity, documentation, and much more generate a myriad of requirements and obligations.

\begin{table}
	\centering
	\caption{Selection of articles of the European Commission's AI legislative proposal \cite{eu_parliament_2024corr} that most likely impact AI-based systems related to automated driving.}
	\begin{tabular}{l p{0.77\linewidth}}
		\toprule
		\multicolumn{2}{c}{\textbf{Chapter III}} \\
		\multicolumn{2}{c}{\textbf{High-risk AI systems}} \\
		\midrule
		\textit{Section 2} & \textit{Requirements for high-risk AI systems} \\
		Article 8 & Compliance with the requirements \\
		Article 9 & Risk management system \\
		Article 10 & Data and data governance \\
		Article 14 & Human oversight \\
		\midrule
		\textit{Section 3} & \textit{Obligations of providers and deployers of high-risk AI systems and other parties} \\
		Article 16 & Obligations of providers of high-risk AI systems \\
		Article 17 & Quality management system \\
		Article 18 & Documentation keeping (new) \\
		Article 23 & Obligations of importers \\
		Article 24 & Obligations of distributors \\
		Article 25 & Responsibilities along the AI value chain \\
		Article 26 & Obligations of deployers of high-risk AI systems \\
		\midrule
		\textit{Section 5} & \textit{Standards, conformity assessment, certificates, registration} \\
		Article 43 & Conformity assessment \\
		Article 44 & Certificates \\
		Article 47 & EU declaration of conformity \\
		\midrule
		\multicolumn{2}{c}{\textbf{Chapter VIII}} \\
		\multicolumn{2}{c}{\textbf{EU database for high risk AI systems}} \\
		\midrule
		Article 71 & EU database for high-risk AI systems listed in Annex III \\
		\midrule
		\multicolumn{2}{c}{\textbf{Chapter IX}} \\ 
		\multicolumn{2}{c}{\textbf{Post-market monitoring, information sharing and  market surveillance}} \\ 
		\midrule
		\textit{Section 1} & \textit{Post-market monitoring} \\
		Article 72 & Post-market monitoring by providers and post-market monitoring plan for high-risk AI systems \\
		\midrule
		\textit{Section 2} & \textit{Sharing of information on serious incidents} \\
		Article 73 & Reporting of serious incidents\\
		\midrule
		\textit{Section 3} & \textit{Enforcement} \\
		Article 76 & Supervision of testing in real world conditions by market surveillance authorities\\
		\bottomrule
	\end{tabular}
	\label{tab:regulation}
\end{table}

In addition to the EU AIA, the Commission has launched an AI innovation package to support start-ups and SMEs in developing trustworthy AI that respects EU values and rules \cite{CommissionAIInnovationPackage2024}. To this end, a Large AI Grand Challenge was launched, offering financial support and access to supercomputers. It was also decided to set up an AI Office within the European Commission. In addition, an amendment to the European High Performance Computing Joint Undertaking Regulation \cite{regulation_1173} was adopted to add a new pillar on the establishment of AI factories. Furthermore, the Commission will financially support generative AI through Horizon Europe and the Digital Europe program and thus expects to attract a total investment of around \euro4 billion by 2027. \cite{CommissionAIInnovationPackage2024}. Beyond that, initiatives such as GenAI4EU or consortiums such as Alliance for Language Technologies \cite{EUDecision2024_458} and CitiVERSE \cite{EUDecision2024_459} have recently been initiated.

\subsection{United States (US)}
In the US, there is no federal legislation governing AI applications. However, in October 2022, the White House Office of Science and Technology Policy (OSTP) published the "Blueprint for an AI Bill of Rights" \cite{blueprintaibillofrights}. This is a guide for development, deployment, and use of AI collaboratively worked out by OSTP, academia, human rights, groups, general public, and large companies (Amazon, Anthropic, Google, Inflection, Meta, Microsoft, and OpenAI). The core of the guideline addresses the following five topics: safe and effective systems; algorithmic discrimination protections; data privacy; notice and explanation; human alternatives, consideration, and fallback. The guide itself is not legally binding. Furthermore, there is a handbook, from principles to practice, which, in combination with the guide, should serve the industry as well as lawmakers at all levels of government who are considering AI regulation. 

In January 2023, the National Institute of Standards and Technology (NIST) published the AI Risk Management Framework \cite{ai2023artificial}. This non-binding framework "...is intended to be voluntary, rights-preserving, non-sector-specific, and use-case independent, and to provide organizations of all sizes, in all sectors, and across society with the flexibility to implement the framework's approaches."\cite{ai2023artificial} The first part of the framework describes risk and trustworthiness in the context of AI in more detail. The second part addresses four key functions for mitigating AI-related risks and creating trustworthy AI systems. This includes establishing governance structures and policies (\textit{govern}), mapping the risks associated with AI systems, including identifying and assessing potential risks (\textit{map}), measuring and assessing the risks associated with AI systems (\textit{measure}) and managing the risks associated with AI systems (\textit{manage}). 

Although no dedicated AI legislation is currently addressed at the federal level, related issues are touched upon by existing laws. For instance, issues related to bias and unfair treatment in AI are in many cases covered by existing laws, e.g., under the Civil Rights Acts of 1964 \cite{CivilRightsAct}, the Equal Credit Opportunity Act (ECOA) \cite{ECOA}, or the Fair Housing Act (FHA) \cite{FairHousingAct}, even though these laws do not specifically relate to AI technology. This shows that common AI issues are often regulated in general terms, which is why a law for unbiased AI, e.g., is not as urgent.
 
In parallel, in single states and cities, governments are pursuing individual regulations. For example, New York City local law 144 \cite{NYCLocalLaw144} requires rules such as a bias audit for the use of AI in the hiring process. Colorado local law SB21-196 \cite{SenateBill21169} prohibits unfair insurance through AI and California local law AB 331 \cite{AB331} requires consequence assessments of AI based decision tools. Beyond that, an inventory of all AI systems require laws currently being drafted in various states like Vermont, Texas, and Washington. In particular, it should be noted that a large number of laws are currently being processed. The following states should be emphasized in this context: California, Illinois, Massachusetts, New Jersey, New York, and Texas. While the aforementioned states predominantly have laws in process akin pending, Texas differs in that two laws are enacted and six are failed-adjourned. Furthermore, several bills have also failed in Maryland, Missouri, and New Mexico. This illustrates that currently there is a considerable amount of movement in this area of legislation and parliamentarians have not yet found a clear path towards the legislative regulation of AI at the state level. A detailed overview of artificial intelligence legislation in 2023 at the state level is provided by the National Conference of State Legislatures \cite{NCSL_AI_2023}.

\begin{table}
	\centering
	\caption{Overview of legal developments in Florida.}
	\begin{tabularx}{\linewidth}{l l X c }
		\toprule
		\textbf{Year} &
		\textbf{Bill} & 
		\textbf{Short Description} & 
		\textbf{Source} \\
		\midrule
		2012&HB 1207& required for testing purposes a human operator in the driver's seat with a valid driver's license to immediately take over active control of the vehicle when necessary.& \cite{fl_hb1207}\\
		2016&HB 7027&extended operation on public roads and reaffirmed the requirement for a valid driver's license but eliminated the requirement of physical presence in the vehicle of the operator.&\cite{fl_hb7027}\\
		2019&HB 311&removed the requirement for a valid driver's license for operating fully autonomous vehicles, as the vehicle is considered the operator.&\cite{florida2019}\\
		\bottomrule
	\end{tabularx}
	\label{tab:florida}
\end{table}
  
In the area of autonomous driving, there are also only a few guidelines at the federal level. Most of these are provided by the National Highway and Transportation Safety Administration (NHTSA). Thus, the regulation of autonomous driving, likewise, mainly takes place at state level and provides different sets of regulations. While a large number of states currently have no regulation (Alaska, Arizona, Delaware, Hawaii, Idaho, Maryland, Massachusetts, Missouri, Montana, New Jersey, New Mexico, Ohio, Rhode Island, Wyoming), other states are making efforts to regulate and are even enacting regulations that weaken or expand previous regulations. An example in this context is Florida, where the legislative development over time with regard to the driver and operator requirements is outlined in Table \ref{tab:florida}. The table highlights in particular the requirements for presence in the vehicle and the need for a driver's license. Similar developments can be observed across other states as well. A selection of bills is provided in Table \ref{tab:all_bills} with respect to operator requirements, driver's license, liability, and specialties. Beyond the table, it can be generally summarized that the increasing development of autonomous driving is accompanied by a more precise definition of terms, requirements, and conditions troughout the legislation. Therefore, it is conceivable that the legal situation responds specifically to the different degrees of technological maturity. The consequence is a far-reaching regulation of diverse aspects based on granular criteria and conditions. 

\begin{table*}
	\centering
	\caption{Overview of the legal situation in a set of US states.}
	\begin{tabularx}{\linewidth}{l l  l l l X c }
		\toprule
		\textbf{State} &
		\textbf{Bill} &
		\textbf{Operator}&
		\textbf{License}&
		\textbf{Liability } &
		\textbf{Specialty} & 
		\textbf{Source} \\
		\midrule
		Alabama&SB 47 (2019)&remote&required&US\$2,000,000&focus on teleoperation systems&\cite{SB47}\\
		California&AB 669 (2017)&tester&required&n/a&focus on testing of vehicle platooning&\cite{AB669}\\
		California&AB 87 (2018)&law enforcement&not required&n/a&focus on removing a vehicle to direct traffic or enforce parking laws and regulations&\cite{AB87}\\
		Connecticut&SB 260 (2017)&in driver’s seat&required&US\$5,000,000&focus on pilot program for the testing purposes&\cite{SB260}\\
		Connecticut&SB 924 (2019)&physically inside&required&US\$5,000,000&Amends Connecticut SB 260&\cite{SB924}\\
		Florida&HB 1207 (2012)&in driver’s seat&required&US\$5,000,000&focus on testing purposes&\cite{fl_hb1207}\\
		Florida&HB 7027 (2016)&not required inside&required&US\$5,000,000&focus on expanding scope and reducing requirements&\cite{fl_hb7027}\\
		Florida&HB 311 (2019)&not required&not required&US\$1,000,000&focus on fully autonomous vehicle&\cite{florida2019}\\
		Georgia&SB 219 (2017)&not required&not required&required&general focus&\cite{georgia_sb219}\\
		Iowa&SF 302 (2019)&not required&required&required&general focus&\cite{ia_sf302}\\
		Louisiana&HB 455 (2019)&remote&required&US\$2,000,000&considers remote and teleoperation&\cite{HB455}\\
		Minnesota&HB 6 (2019)&in driver’s seat&required&required&focus on platooning, requirements hold for each vehicle in the platoon&\cite{M_HB_6}\\
		Nebraska&LB 989 (2018)&not required&depends&required&general focus&\cite{n_LB_989}\\
		Nevada&AB 511 (2011)&n/a&endorsement&required&general focus&\cite{AB511}\\
		Nevada&AB 69 (2017)&not required&not required&US\$5,000,000&general focus&\cite{AB69_N}\\
		New Hampshire&SB 216 (2019)&not required&required for tests&US\$5,000,000&with a focus on testing&\cite{nh_sb216_fn}\\
		North Carolina&HB 469 (2017)&n/a&not required&required&focus on fully autonomous vehicle&\cite{nc2017sessionlaw}\\
		North Dakota&HB 1418 (2019)&not required&not required&n/a&general focus&\\
		Texas&SB 2205 (2017)&not required&not required&required&general focus&\cite{SB2205}\\
		\bottomrule
	\end{tabularx}
	\label{tab:all_bills}
\end{table*}

Another key aspect that can be observed throughout the legislation is the specific consideration of testing. This is particularly relevant for development and is often subject to specific requirements. Without going into any detail, a rough insight is provided in the following due to its relevance for research and development. For instance, New York requires state police oversight and reporting for testings and establishes requirements for demonstrations and operations \cite{nya09508}. Apart from New York, Vermont's legislation also creates a solid basis for the careful approval of test vehicles \cite{vt_s0149}. For example, it defines a comprehensive list of criteria and requirements. Beyond that, local authorities, e.g. the transport authority, are given the right to adopt and introduce new rules. Furthermore, Colorado and Iowa also have more specific conditions and requirements \cite{co_sb17_213, co_sb19_239, ia_sf302}. In addition to those mentioned previous, California should be emphasiszed in this context, as its legislation on testing and operation \cite{ca_sb1298, ca_ab1592, ca_ab669, ca_ab1444} is far-reaching and has constantly evolved over time due to the local presence of the tech industry.   In addition to all these primarily important regulations, bills have also been passed that go beyond this, such as the taxation of autonomous vehicles \cite{ca_ab1184}.

\begin{figure}
	\centering	
	\includegraphics[width=\linewidth]{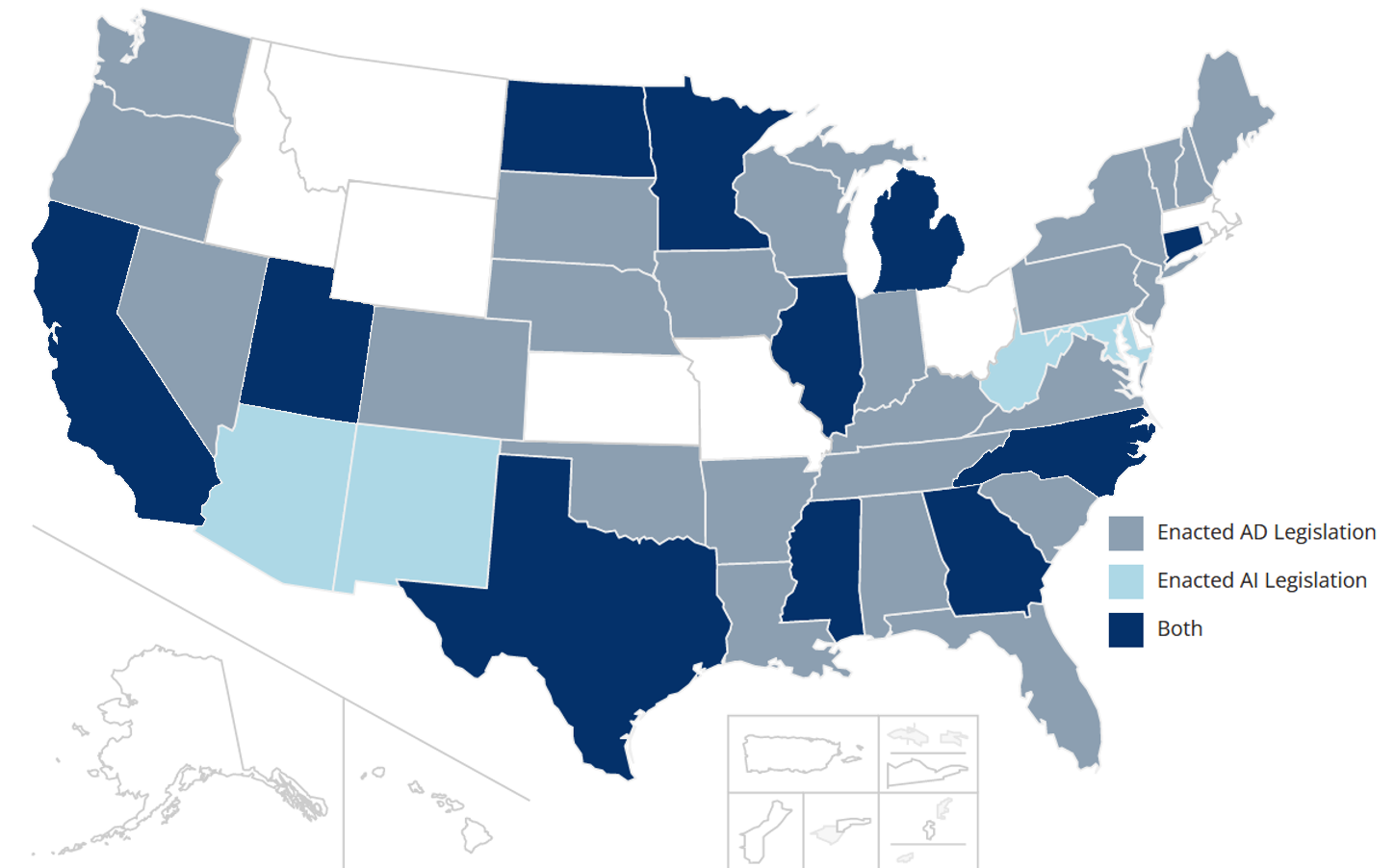}
	\caption{Visualization of the legal regulation of artificial intelligence and autonomous driving in the United States as of 23/04/2024, inspired by \cite{NCSL_AD_2020}.}
	\label{fig:usa_rules}
\end{figure}

It can therefore be concluded that a bottom-up approach is being pursued in the US in the regulation of both artificial intelligence and autonomous driving. An overview of the current legal situation, the enacted legislation, is summarized in Figure \ref{fig:usa_rules}. 

\subsection{China}
In China, the "Action Outline for Promoting the Development of Big Data" was published in 2015 \cite{bigdatadevelopmentplan}. In July 2017, the "Next Generation Artificial Intelligence Development Plan" was issued by the State Council \cite{statecouncil2017}. The plan focuses on promoting AI development, but also includes a timeline for developing AI governance regulations by 2030. Along, China has set a goal of generating US\$154 billion in annual revenue by 2023 and becoming the global AI leader by 2030 in AI industry. 

In June 2019, the "Governance Principles for New Generation AI: Develop Responsible Artificial Intelligence" was published by the National New Generation AI Governance Expert Committee, which considers eight principles (harmony, friendliness, fairness, inclusiveness, respect for privacy, security and controllability, shared responsibility, open collaboration, and agile governance) for AI governance \cite{aigovernanceprinciples}. In December 2020, recommendation algorithms were specifically addressed in the Party document "Outline for Establishing a Rule-of-Law-Based Society (2020-2025)" \cite{ccplegaloutline2020} with regards to social and legal concerns.

In September 2021, the "Guiding Opinions on Strengthening Overall Governance of Internet Information Service Algorithms" \cite{internetalgorithmgovernance} and "Ethical Norms for New Generation AI" \cite{aiethicsguidelines} were published by the Cyberspace Administration of China (CAC) and the National New Generation AI Governance Expert Committee, respectively. The former was co-signed by many bodies and provides general guidelines for the regulation of online algorithms through 2024, while the latter provides comprehensive guidance on ethical norms that should be considered in AI governance. 

In December 2021, the first binding regulation on algorithms was published by the CAC and others under the title "Provisions on the Management of Algorithmic Recommendations in Internet Information Services" \cite{internetalgorithmregulations}. The regulation centrally addresses 
cybersecurity concerns and regulates algorithms for online distribution of content and news. 
In March 2022, the Regulation "Opinions on Strengthening the Ethical Governance of Science and Technology" was published by the CPC Central Committee, State Council \cite{technologyethicsgovernance}. It aims to deploy ethics and governance in research and development. AI is one of three key aspects within. 
In November 2022, "Provisions on the Administration of Deep Synthesis Internet Information Services" was published by the CAA among others \cite{internetdeepsynthesisregulations}. This regulation focuses on AI applications for text, audio, and video generation, and specifically prohibits "fake news".

In April 2023, CAC responded to the development of Generative AI, such as ChatGPT, by publishing a first draft of "Measures for the Management of Generative Artificial Intelligence Services" \cite{aiserviceregulations}. This regulation covers similar areas as the deep synthesis regulation, but the emphasis on data, especially training data, is considerably stronger. Moreover, providers of such systems are required to ensure that the content generated is true and accurate.

At the city level, the "Shanghai Regulations on Promoting the Development of the AI Industry" took effect in October 2022 \cite{shanghaiairegulation}. The focus here is on increasing the accumulation of AI industry by specifically promoting innovation and achieving breakthroughs. Dedicated management combined with "sandbox" oversight is intended to create the necessary space for research, development and testing of AI systems. Regulation in Shanghai even goes so far as to tolerate minor rule violations. This is in recognition of the fact that AI research involves a number of unanswered questions and scientific exploration of frontiers and discovery of breakthroughs should not be hindered by regulations. At the same time, the regulation requires the Shanghai municipality to establish an AI ethics expert committee to set principles and ethical standards. 

In addition, the Shenzhen regulation "Regulations for the Promotion of the Artificial Intelligence Industry in Shenzhen Special Economic Zone" came into effect in November 2022 \cite{shenzhenaipromotionregulation}. In particular, this regulation aims to promote, enable, and encourage local industry to take the lead in AI research. The regulation also provides guidelines for data sharing between organizations and companies. Notably, the regulation provides for resident AI systems that are deemed low-risk to be more easily tested and trialed. In this context, Shenzhen is also establishing an AI ethics council to develop safety standards and investigate the technology impact. To achieve the goal of a pivotal AI leadership, Shenzhen invests over US\$108 billion from 2021 to 2025 \cite{ChinaScience2021}. 

Shenzhen is also pushing ahead with regulation in the area of autonomous driving \cite{shenzhenautonomousdrivingregulation}. In large parts of the city, licensed vehicles are allowed to drive without a driver in the driver's seat, in case a supervisor is in the vehicle. It is also worth noting that a framework for liability \cite{kirton2022shenzhen} in driving-related accidents has been introduced in this context. 

\subsection{Further Selected National Aspirations}
In addition to the key regions of the EU, US and China, this section also briefly introduces Canada, Japan and Switzerland. Canada places a stronger emphasis on data in its regulatory framework, Japan prioritizes societal impact and Switzerland advocates for further research. All of these countries provide valuable insights and impulses that are worth to be considered.

In 2017, \textit{Canada} launched a nationwide AI strategy, the "Pan-Canadian Artificial Intelligence Strategy", which is aimed at commercialization, standardization, research, and the fostering of talents \cite{pancanadianaistrategy}. Moreover, the Canadian Parliament tabled a draft Artificial Intelligence and Data Act (AIDA) in June 2022 that emphasizes a risk-based approach \cite{billc27}. While the EU includes bans in case of unacceptable risks, this draft does not consider bans. However, in high-risk applications, a mitigation plan is required to reduce risk and increase transparency.

\textit{Japan}'s "AI Strategy 2022" is closely linked to the "Society 5.0" project \cite{cabinetofficeaistrategy2022}. In addition, the "Social Principles of Human-Human-Centric AI" \cite{socialprinciplesai} presented by the Integrated Innovation Strategy Promotion Council deal, on the one hand, with key social principles that should be taken into account in the context of AI. On the other hand, the publication contains guidelines for AI development. In July 2021, "Governance Guidelines for Implementation of AI Principles" \cite{metiaigovernanceguidelines} were also published to support developers and refer to various national and international guidelines. However, all these publications are not legally binding regulatory laws.

In \textit{Switzerland}, a position paper of the Digital Society Initiative on a legal framework for artificial intelligence was published in November 2021 \cite{dsiaiframework}. The need for regulation is acknowledged; passive adoption of the EU regulation should not take place, instead Switzerland should develop its own regulations and standards. The position paper particularly emphasizes that regulation must fulfill two objectives. On the one hand, sufficient freedom should be left for the development and use of AI. On the other hand, the use of AI must not disadvantage anyone and/or society or undermine principles of the rule of law. The position paper also argues less for a general AI law and more for the adaptation of existing laws and the development of sector-specific standards. In addition, a solid scientific basis should be created by intensifying relevant research. An interdisciplinary expert commission shall be set up to develop standards and regulations.

\subsection{Discussion}
This overview of the current regulation of AI systems and the focus on autonomous vehicles demonstrates that a need for regulation of AI systems is recognized. Furthermore, it illustrates that a variety of different regulation approaches coexist. In particular, the regulation approaches vary between strict regulation with high safety requirements, which may limit the use of the technology, and lax regulation aimed at exploring the technology at the frontier in order to achieve technological leadership.

Due to the rapid development of AI, the recurring breakthroughs and the multitude of unanswered questions, regulation tends to fall short. Uncertainty about the impact of regulation on research and development and the market in general often leads to legally non-binding guidelines. This is permissible as key aspects of fundamental rights are enshrined in the respective national legislation, which counteracts the misuse of AI technologies to a certain extent. 

Within regulation, the central importance of data is increasingly being taken into account and specifically addressed. For instance, several countries introduced regulations for generative AI models based on training data to improve transparency. Also noteworthy is the significant consideration of data throughout the entire AI life cycle in the EU AIA. Overall, this demonstrates the enhanced awareness of the importance of data for AI legislation. In the field of autonomous driving, there is an increasing awareness of the need to test and evaluate systems on real roads under real conditions, even if full validation is not yet possible. However, it is also apparent that there are very different approaches to the testing, approval and use of robotaxis, for example. In general, the legal and regulatory landscape is complicated and changing due to the rapid development of technology. Note that this overview of legislation is based on an analysis up to 23/04/2024.

In our assessment, it is particularly important that the concrete definition of risks will be addressed more specifically in the course of regulation in the future. From a technical perspective, it is obvious with systems as complex and emergent as autonomous driving that zero percent risk is not feasible. Some residual risk will always remain. Especially the definition of a socially acceptable risk is indispensable on the long run. Particularly in the case of autonomous driving, the close connection between acceptable risk and ethical concerns becomes apparent. While technology research can identify functional possibilities and technical limits, regulation is needed to define acceptable risk levels, beyond a mortality rate. We believe it is appropriate to involve society in this process in order to communicate the persistence of a residual risk and to gain societal acceptance in the process of definition.  
\section{Open Questions \& Potential Solutions}\label{sol}
Having explored the research, standardization, and regulation of AI safety assurance for automated vehicles in the previous sections, this section aims to present the key open questions for future research and potential solutions to provide a forward-looking perspective on the field. 

Overall, it is clear that a shift in the field of safety assurance is likely to be as disruptive as the technology of AI itself to adequately address the challenges. In particular, a shift from the mathematically explicit to the data-based implicit seems to be unfolding. While this seems reasonable and practicable at first glance, it raises numerous research questions. However, the transition from classic SIL/ASIL to AI-SIL is a first positive example of a transformation towards the future approaches. Likewise, it is conceivable that functional safety and SOTIF could be transformed into a data-based functional or SOTIF safety assurance. However, on closer examination, it remains valid to raise the question of how a data-based analysis of AI systems can be conducted so that reliable statements about safety can be made. In the field of systems theory, however, no general breakthrough in this direction is yet apparent. It should also be noted that AI systems can be very heterogeneous, ranging from machine learning to deep learning, reinforcement learning, and generative AI systems. Furthermore, there are many other distinguishing criteria. Accordingly, the open research questions and possible solutions appear to be very wide-ranging when going deeper. Thus, it opens up a broad spectrum of interesting possible future research opportunities. 

In addition to the in-depth research questions and possible research directions mentioned so far, the general safety assurance process is also a key challenge. Due to rapidly changing environments, it may be necessary to regularly update and release AI systems. Accordingly, on a more abstract level, an iterative, data-based development and verification and validation process generally seems to be a possible solution. 
\section{Summary \& Conclusion}\label{SafetySummary_chapter5}
There are still many unanswered questions in the field of AI assurance research. Clearly, the increasingly complex architectures and emerging approaches and methods present particular challenges. Notably, the immediate dependence of systems on data as well as the implicit nature of representing non-trivial relationships highlight the need for new approaches--with data being pivotal. The extension of familiar assurance methods, such as functional assurance to data-based functional assurance, as well as the transition from rule-based to data-based assurance are necessary in our perspective. 
The importance of assurance agnosticism towards hardware and new methods and approaches is well known in AI research and autonomous driving research and related standardization. Both standardization and regulation address the distinct relevance of taking a holistic view across the entire lifecycle. While data also holds a special role here, it has not yet been placed at the center of lifecycle assurance. Therefore, future work should focus on data-driven functional safety and repetitive exploration, observation and mitigation throughout the lifecycle.

From a holistic perspective, we argue for stronger networking of research, standardization, and regulation as well as for more flexibility in standardization and regulation in order to keep pace with the rapid development in order to not fall too far behind. However, to avoid too strong impact and indirect control of indispensable research and development, we consider open innovation and closely collaboration-driven open standards and regulations that are not legally binding to be useful. The non-legally binding way can thus serve as a guideline, which does not have to be followed compulsorily. This openness allows and encourages new ways to be considered and, if successful, to be introduced in turn, thereby ensuring continuous improvement. This would allow good ideas and approaches to be processed across national borders, legal systems, and governmental systems, which could have a long-term impact on local, national, and international legislation. Thus, AI safety assurance empowers using AI for the benefit of humanity and to prevent harm.

\bibliographystyle{IEEEtran}
\bibliography{literature}
\vspace{-10 mm}
\begin{IEEEbiography}[{\includegraphics[width=1in,height=1.25in,clip,keepaspectratio]{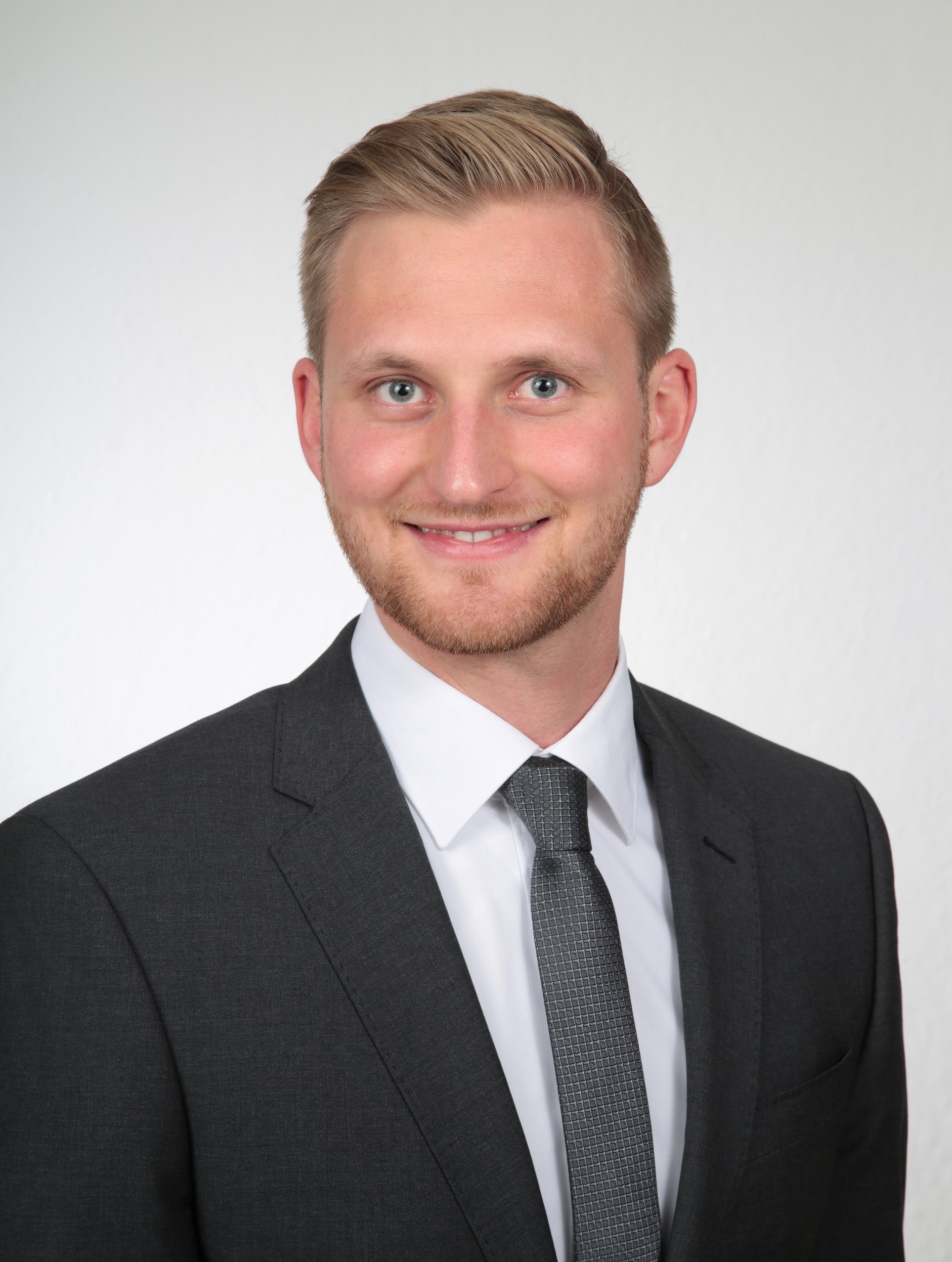}}]{Lars Ullrich}
	received the M.Sc. degree in mechatronics from Friedrich–Alexander–Universita\"at Erlangen–N\"urnberg, Germany, in 2022, where he is currently pursuing the Ph.D.
	(Dr.Ing.) degree with the Chair of Automatic Control.
	
	His research interests include probabilistic trajectory planning for safe and reliable autonomous driving in uncertain dynamic environments with a focus on addressing challenges arising from the use of AI systems in automated driving.
\end{IEEEbiography} \vspace{-10 mm}
\begin{IEEEbiography}[{\includegraphics[width=1in,height=1.25in,clip,keepaspectratio]{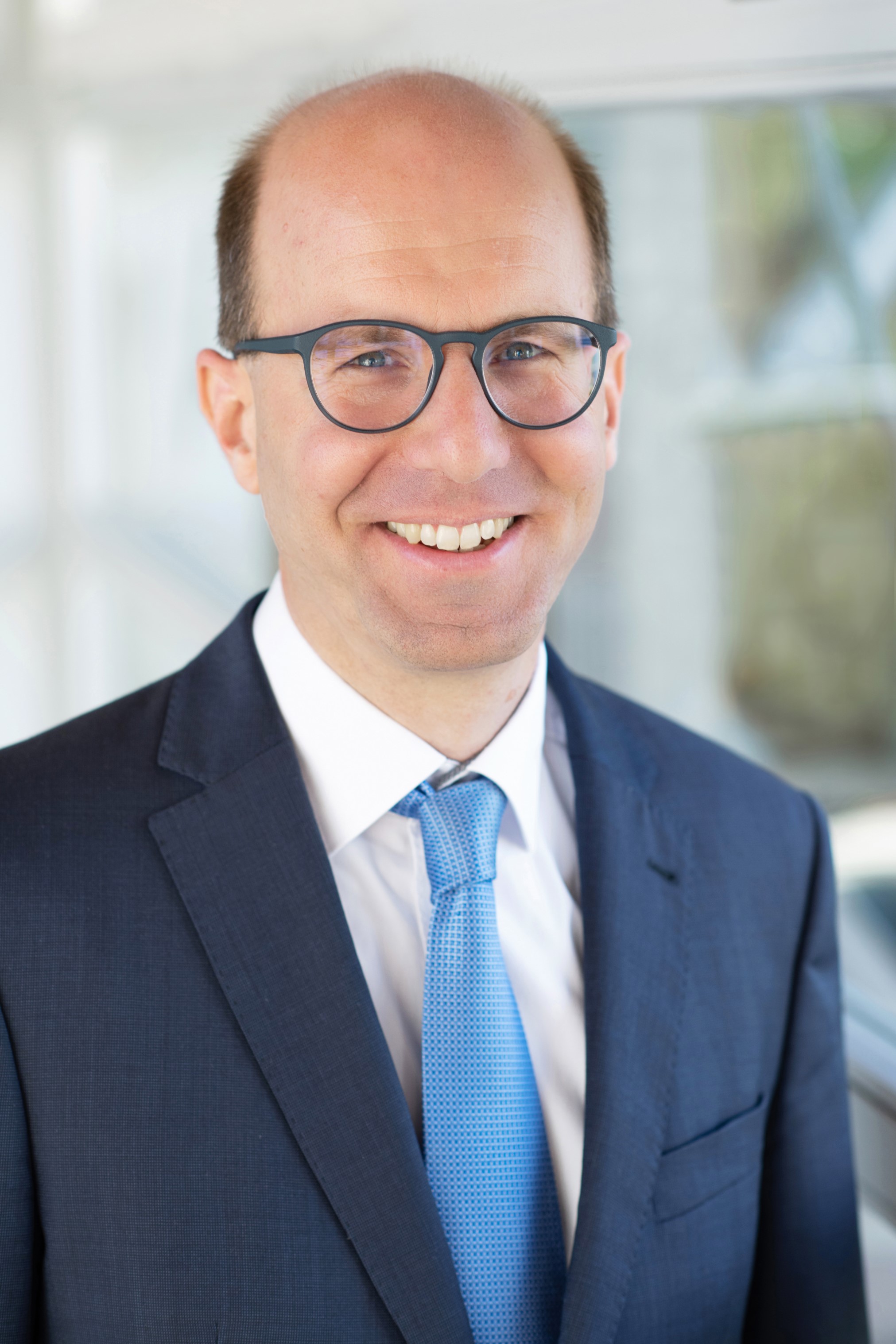}}]{Michael Buchholz}
	received his Diploma degree in Electrical Engineering and Information Technology as well as his Ph.D. from the faculty of Electrical Engineering and Information Technology at University of Karlsruhe (TH)/Karlsruhe Institute of Technology, Germany.  He is a research group leader and lecturer at the Institute of Measurement, Control, and Microtechnology, Ulm University, Ulm, 89081, Germany, where he earned his ``Habilitation'' (post-doctoral lecturing qualification) for Automation Technology in 2022. His research interests comprise connected automated driving, electric mobility, modelling and control of mechatronic systems, and system identification.
\end{IEEEbiography} \vspace{-10 mm}
\begin{IEEEbiography}[{\includegraphics[width=1in,height=1.25in,clip,keepaspectratio]{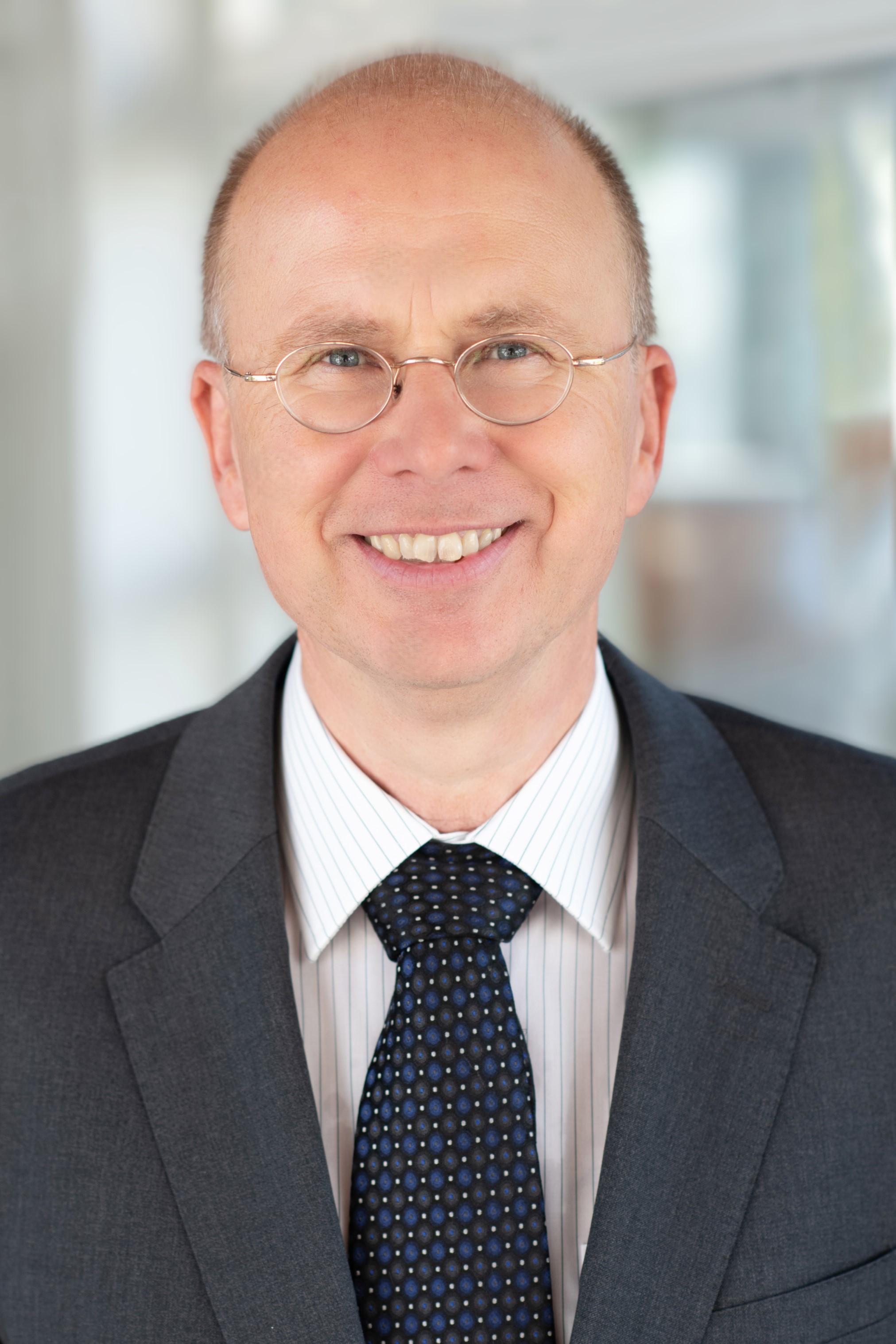}}]{Klaus Dietmayer}
	(Senior Member, IEEE) earned his degree in electrical engineering from the Technical University of Braunschweig, Germany and completed his Ph.D. in 1994 at the University of the Armed Forces, Hamburg, Germany. Afterwards, he began his industrial career as a research engineer at Philips Semiconductors, Hamburg, progressing through various roles to become the manager for sensors and actuators in the automotive electronics division. 
	
	In 2000, Dietmayer was appointed as a Professor of Measurement and Control at the University of Ulm. He currently serves as the Director of the Institute for Measurement, Control, and Microtechnology within the School of Engineering and Computer Science.
	
	His primary research interests include information fusion, multi-object tracking, environment perception, situation assessment, and behavior planning for autonomous driving. The institute operates three automated test vehicles with special licenses for public road traffic, along with a test intersection equipped with infrastructure sensors for evaluating automated and networked cooperative driving in Ulm.
\end{IEEEbiography} \vspace{-10 mm}
\begin{IEEEbiography}[{\includegraphics[width=1in,height=1.25in,clip,keepaspectratio]{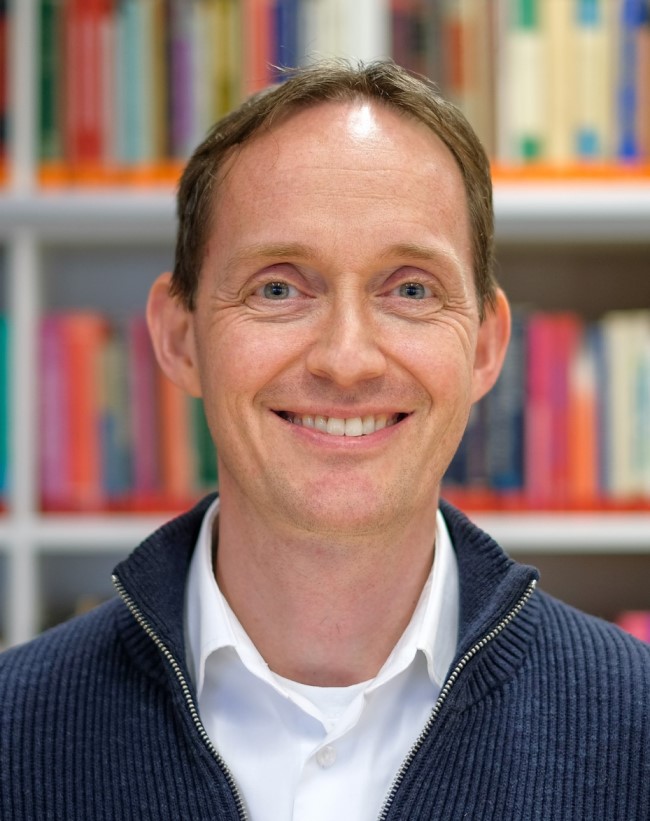}}]{Knut Graichen}
	(Senior Member, IEEE) received the Diploma-Ing. degree in engineering cybernetics and the Ph.D. (Dr.-Ing.) degree from the University of Stuttgart, Stuttgart, Germany, in 2002 and 2006, respectively.
	
	In 2007, he was a Post-Doctoral Researcher with the Center Automatique et Syst\`emes, MINES ParisTech, France. In 2008, he joined the Automation and Control Institute, Vienna University of Technology, Vienna, Austria, as a Senior Researcher. In 2010, he became a Professor with the Institute of Measurement, Control and Microtechnology, Ulm University, Ulm, Germany. Since 2019, he has been the Head of the Chair of Automatic Control, Friedrich–Alexander–Universita\"at Erlangen–N\"urnberg, Germany. His current research interests include distributed and learning control and model predictive control of dynamical systems for automotive, mechatronic, and robotic applications.
	
	Dr. Graichen is the Editor-in-Chief of Control Engineering Practice.
\end{IEEEbiography}

\end{document}